\documentclass[12pt, a4paper]{article}
\pdfoutput=1

\usepackage{amsmath}
\usepackage{amsfonts}
\usepackage{amssymb}
\usepackage{latexsym}
\usepackage{graphicx}
\usepackage{color}
\usepackage{mathrsfs}
\usepackage{slashed}
\usepackage{hyperref}
\usepackage{datetime}
\usepackage{enumerate}

\numberwithin{equation}{section}

\hypersetup{colorlinks, citecolor=bluscuro, linkcolor=bluscuro, urlcolor=bluscuro}
\definecolor{rossos}{rgb}{0.8,0.2,0.3}
\definecolor{bluscuro}{rgb}{0.15, 0.2, 0.9}
\definecolor{verdes}{rgb}{0.1, 0.5, 0.1}

\makeatother   

 \def\be   {\begin{equation}}   \def\ee   {\end{equation}}
 \def\ba   {\begin{array}}      \def\ea   {\end{array}}
 \def\bea  {\begin{eqnarray}}   \def\eea  {\end{eqnarray}}
 \def\bean {\begin{eqnarray*}}  \def\eean {\end{eqnarray*}}

\begin{document}


\vspace{7cm}

\begin{center}
\vspace{5cm}
{\LARGE \textsc {
What the top asymmetries tell us about\\ [0.5 cm]
single top production and Higgs decays
}}
\\ [2.5cm]
{\large{\textsc{
Wei-Chih Huang$^{\,a,b}$ and Alfredo Urbano$^{\,a}$}}}
\\[1cm]

\large{\textit{
$^{a}$~SISSA, via Bonomea 265, I-34136 Trieste, ITALY.\\ \vspace{1.5mm}
$^{b}$~INFN, sezione di Trieste, I-34136 Trieste, ITALY.
}}
\\ [3.5cm]
{ \large{\textsc{
Abstract
}}}
\\ [1.5cm]
\end{center}

The top asymmetries measured at the Tevatron, and the discrepancy in the di-photon Higgs rate under investigation at the LHC represent two possible hints of  physics beyond the Standard Model (SM). In this paper we address the possibility to explain and relate both these experimental anomalies with the help of an extra scalar doublet in addition to the SM particle spectrum. The connection is provided by the constraints that the electroweak precision tests impose on the oblique S and T parameters.\\ As a result, considering the semi-leptonic single top production in the tW-channel at the LHC, our analysis \textit{predicts} a bump in the invariant mass distribution of the two light jets.


\def\thefootnote{\arabic{footnote}}
\setcounter{footnote}{0}
\pagestyle{empty}

\newpage
\pagestyle{plain}
\setcounter{page}{1}

\section{Introduction}
The ElectroWeak Symmetry Breaking (EWSB), one of the most profound enigmas in theoretical high-energy physics, is close to be solved. Nevertheless, even after the discovery at the LHC  \cite{Giannotti,:2012gk,Incandela,:2012gu} of a particle similar to the Higgs boson \cite{Higgs} in the Standard Model (SM), Nature has chosen to keep us with bated breath to the last. The true aspect of the Higgs boson, in fact, is still shrouded in mystery, and only a careful study of its couplings will tell us if any New Physics (NP) is really hidden under its SM appearance.\\
On the other hand, even more than 17 years after its discovery at the Tevatron in 1995 \cite{Abachi:1995iq}, the role of the top quark in the SM is still in some respects unknown. In particular the most intriguing propriety is the mass. With a value around $173$ GeV \cite{Amsler:2008zzb} - more than 35 times larger than that of its iso-partner, the bottom quark -  it is extremely close to the electroweak scale, and for this reason the top quark is the only fermion in the SM for which the coupling with the Higgs boson is sizable. This fact suggests that the top quark might play a crucial role in the realization of the EWSB mechanism.\\
With this perspective in mind, in this paper we try to analyze two recent experimental anomalies. On the one hand at the LHC, the biggest deviation from the SM associated with the Higgs boson is the di-photon decay channel, whose branching ratio seems to be around two times larger with respect to the correspondent SM value. In particular ATLAS reports a signal strength of $1.8\pm 0.5$ times the SM value \cite{ATLAShcps}, while the observed strength at CMS is $1.56\pm 0.43$ \cite{CMShcps}.
 This loop-induced decay channel is particularly sensitive to any NP contributions, and a large number of interesting theoretical explanations for this excess have been pursued involving new scalars \cite{calderonespin0}, fermions \cite{calderonespin1/2}, gauge bosons \cite{calderonespin1}, or spin-$2$ particles \cite{calderonespin2}. On the other side of the ocean both CDF and D$\emptyset$ collaborations have measured the Forward-Backward (FB) asymmetry of the top quark \cite{Aaltonen:2011kc,Aaltonen:Note,Abazov:2011rq}. Their experimental results seem to point towards a mild excess with respect to the SM prediction. In particular both CDF \cite{Aaltonen:Note}  and D$\emptyset$ \cite{Abazov:2011rq} report a $\sim 1.5\,\sigma$ deviation from the SM considering the inclusive FB asymmetry. For large values of invariant mass of the top-antitop pair, on the contrary, CDF reports a  $\sim 2.4\,\sigma$ deviation \cite{Aaltonen:Note}. If confirmed, this discrepancy might suggest the presence of NP affecting the top quark pair production, and several models have been proposed. These models fall into two main categories;
 s-channel exchange of vector mediators with axial couplings \cite{axigluon} or t-channel exchange of flavor-violating mediators, either scalars \cite{FBscalars} or vectors \cite{FBvectors}.\\
 In this paper we focus on the possibility to explain and relate both these experimental anomalies with the help of an extra scalar doublet in addition to the SM. The neutral component of the doublet mediates the t-channel top pair production generating the required FB asymmetry while its charged component enters in the loop-induced decays $h\to \gamma\gamma$, $h\to Z\gamma$. Moreover, we analyze the consequences of this phenomenological scenario on the single top production at the LHC. This paper is organized as follows. In Section \ref{sec:ScalarDoublet} we set the ground for our analysis by introducing, following  \cite{Blum:2011fa}, the extra scalar doublet and specifying its Yukawa couplings with the SM fermions. Section \ref{sec:TopAsymmetries} is dedicated to the top asymmetries, and in particular we consider the latest experimental results both at the Tevatron and the LHC. In Section \ref{sec:SingleTopProduction} we explore the role of the scalar doublet in the semi-leptonic single top production in the tW-channel at the LHC. We find that it appears as a bump in the invariant mass distribution of the two light jets. This prediction is the most important result of this work. In Section \ref{sec:AFB} we study the effect of the scalar doublet on the radiative Higgs decays $h\to \gamma\gamma$ and $h\to Z\gamma$. We stress that the connection with the FB asymmetry in Section \ref{sec:TopAsymmetries} is provided by the constraints from the electroweak precision tests via the oblique S and T parameters. Finally, we conclude in Section \ref{sec:Conclusions}.

\section{Scalar mediated top asymmetries}
\label{sec:ScalarDoublet}
Following \cite{Blum:2011fa}, we introduce the extra scalar doublet
\begin{equation}\label{eq:ScalarDoublet}
\phi=
\left(
\begin{array}{c}
 \phi^0  \\
   \phi^{-}
\end{array}
\right)~,~~~~~\phi\sim (\textbf{1},\textbf{2})_{-1/2}~,
\end{equation}
where the gauge transformation follows the notation $({\rm SU}(3)_{\rm C},{\rm SU}(2)_{\rm L})_{{\rm U}(1)_{\rm Y}}$.
From the flavor constraints derived in \cite{Blum:2011fa} and in order to account for the top quark FB asymmetry measured at the Tevatron, we constrain ourself to Yukawa couplings with the right-handed up-type quarks only.
As a result, our phenomenological Lagrangian in the mass eigenstates takes the following form
\begin{equation}\label{eq:PhenoLagrangian}
\mathcal{L}=
2\lambda\phi^0
\sum_{q=u,c,t}\overline{q}V_{qb}P_Ru
+2\lambda\overline{b}\phi^-P_Ru
+h.c.~,
\end{equation}
where $\lambda$ is a complex coupling while $V_{qb}$ are the elements in the third column of the Cabibbo-Kobayashi-Maskawa (CKM) matrix \cite{Amsler:2008zzb,CKM}. As shown in Appendix \ref{app:MirrorSymmetry}, the mirror symmetry is able to accommodate the new scalar doublet but can \textit{not} explain the special flavor structure required for the top quark FB asymmetry.

\section{The top asymmetries}
\label{sec:TopAsymmetries}

In this Section, we focus on the top asymmetries and the related leptonic asymmetries in the presence of the scalar doublet $\phi$. Considering the $t\overline{t}$ pair production, these asymmetries are generated by the t-channel exchange of its neutral component $\phi^0$ \cite{Blum:2011fa}. Section \ref{sec:FBTevatron} is devoted to the FB asymmetries measured at the Tevatron while in Section \ref{sec:FBLHC} we discuss the charge asymmetries at the LHC. Note that in this Section we try to reproduce the top asymmetries by focusing on small values for the mass $m_0$ of the neutral component $\phi^0$. As we shall see in Section \ref{sec:AFB}, in fact, this issue will play a crucial role on relating the top asymmetries to the excess of $H\rightarrow \gamma \gamma$ at the LHC.
\subsection{The top and leptonic forward-backward asymmetries at the Tevatron}
\label{sec:FBTevatron}
 In terms of the top-quark pair production at hadron colliders, the charge asymmetry is defined as,
\begin{equation}\label{eq:Cgeneraldefinition}
A_{\rm C}\equiv \frac{N_t(p)-N_{\overline{t}}(p)}{N_t(p)+N_{\overline{t}}(p)}~,
\end{equation}
where $N_i(j)$ represents the number of particles $i$ in directions of particle $j$. At the
Tevatron, this charge asymmetry is equivalent to a FB asymmetry
\begin{equation}\label{eq:FBgeneraldefinition}
A_{\rm FB}\equiv \frac{N_t(p)-N_{t}(\overline{p})}{N_t(p)+N_{t}(\overline{p})}~,
\end{equation}
because $N_{\overline{t}}(p)=N_{t}(\overline{p})$ if CP invariance is good. In the SM \cite{Kuhn,Hollik}, considering a center-of-mass energy $\sqrt{s}=1.96$ TeV for a $p\overline{p}$ collision, the pair production of top quarks proceeds mainly via strong interactions at $\mathcal{O}(\alpha_s^2)$, and in particular the quark-antiquark annihilation process $q\overline{q}\to t\overline{t}$ dominates ($85\%$) with respect to the gluon fusion $gg\to t\overline{t}$ ($15\%$).
These tree level processes almost determine integrated cross section, i.e. the denominator in Eq. (\ref{eq:FBgeneraldefinition}). On the other hand their contribution to the FB asymmetry is zero, by virtue of QCD vector-type interactions. It follows that the forward-backward (FB) asymmetry in Eq. (\ref{eq:FBgeneraldefinition}) receives contributions from \emph{i}) $\mathcal{O}(\alpha_s^3)$ radiative corrections to quark-antiquark annihilation $q\overline{q}\to t\overline{t}$, $q\overline{q}\to t\overline{t}g$ and interference between different amplitudes contributing to gluon-quark scattering $gq\to t\overline{t}q$, $g\overline{q}\to t\overline{t}\overline{q}$, \textit{ii})  $\mathcal{O}(\alpha^2)$ tree level EW quark-antiquark annihilation processes mediated by $\gamma,Z$ exchange, and  \textit{iii}) $\mathcal{O}(\alpha\alpha_s^2)$ EW-QCD interference effects. On the other hand, the scalar doublet $\phi$ contributes to both the asymmetry and top-quark pair production via the tree level annihilation process $u\overline{u}\to t\overline{t}$ through the t-channel exchange of the neutral component, specified in Eq. (\ref{eq:PhenoLagrangian}). Next, we will evaluate this asymmetry in light of the latest experimental results from CDF and D$\emptyset$ collaborations. After a more detailed introduction in Sections \ref{subsec:TevatronAsymmetry} and \ref{subsec:OtherConstraints}, we show our results in Section \ref{subsec:TevatronResults}.

\subsubsection{Description of the asymmetries}\label{subsec:TevatronAsymmetry}

For top-quark pair production at the Tevatron
\begin{equation}\label{eq:TevatronProcessA}
p\,\overline{p}\to t\,\overline{t},
\end{equation}
 the top FB asymmetry in the $t\overline{t}$-rest frame and in a particular bin of invariant mass $M_{t\overline{t},i}$ is defined as
\begin{equation}\label{eq:FBdefinition}
A_{\rm FB}^{t\overline{t}}(M_{t\overline{t},i})\equiv
\frac{N(\Delta y>0,M_{t\overline{t},i})-N(\Delta y<0,M_{t\overline{t},i})}
{N(\Delta y>0,M_{t\overline{t},i})+N(\Delta y<0,M_{t\overline{t},i})}~,
\end{equation}
where $\Delta y\equiv y_t-y_{\overline{t}}$ is the difference between the rapidity of the top and anti-top. Notice that in Eq. (\ref{eq:FBdefinition}) the difference $\Delta y$ is invariant  under a boost transformation along the beam axis, and therefore $A_{\rm FB}^{t\overline{t}}$ in the $t\overline{t}$-rest frame is equal to that in the hadronic (laboratory) frame in the limit of the zero transverse momentum.
On the other hand, considering the production of the $t\overline{t}$-pair as an intermediate state, one can study the following di-leptonic final state
\begin{equation}\label{eq:TevatronProcessB}
p\,\overline{p}\to t\,\overline{t}\to l^{+}\,l^{-}\,j_{b}\,j_{\overline{b}},
\end{equation}
where $l=e,\,\mu$ and $j_{b,\overline{b}}$ denotes a b-tagged jet. In close analogy with Eq. (\ref{eq:FBdefinition}), one can define the inclusive leptonic pair asymmetry,
\begin{equation}\label{eq:LeptonicFB}
A_{\rm FB}^{ll}\equiv
\frac{N(\Delta \eta_l>0)-N(\Delta \eta_l<0)}
{N(\Delta \eta_l>0)+N(\Delta \eta_l<0)}~,
\end{equation}
where $\Delta \eta_l\equiv \eta_{l^+}-\eta_{l^{-}}$ is the difference between the pseudo-rapidity of the dileptonic pair.\\
We summarize the experimental results from both CDF and D$\emptyset$ collaborations in Table \ref{tab:TevatronData}.
These results refer to an unfolded analysis, in which the effects of the detector resolution and acceptance are subtracted. Thus, throughout this Section, we do not attempt to simulate parton showering, hadronization, or detector reconstruction but focus on the top asymmetries at the parton level.
\begin{table}
\begin{center}
\begin{tabular}{ | c | c | c | c | c |}
\hline
  & ${\rm CDF}$\cite{Aaltonen:2011kc}  & ${\rm CDF}$\cite{Aaltonen:Note} &  ${\rm D}\emptyset$\cite{Abazov:2011rq} & ${\rm SM}$\cite{Bernreuther:2012sx} \\
  \hline
 $A_{\rm FB}^{t\overline{t}}({\rm inclusive})$  & $0.158\pm 0.075$  & $0.162\pm 0.047$ & $0.196\pm 0.065$ & $0.088\pm 0.006$ \\ \hline
 $A_{\rm FB}^{t\overline{t}}(M_{t\overline{t}}> 450\,{\rm GeV})$  & $0.475\pm 0.114$   & $0.296\pm0.067$  & $\ast$ & $0.129^{+0.008}_{-0.006}$ \\ \hline
  $A_{\rm FB}^{t\overline{t}}(M_{t\overline{t}}\leqslant 450\,{\rm GeV})$ &  $-0.116\pm 0.153$ & $0.078\pm 0.054$ & $\ast$ & $0.062^{+0.004}_{-0.003}$ \\ \hline
 $A_{\rm FB}^{ll}$  & $\ast$ & $\ast$ & $0.053\pm 0.084$ & $0.048\pm0.004$ \\
  \hline
\end{tabular}
\end{center}
\caption{\emph{Unfolded experimental results from both CDF and D$\emptyset$ collaborations for the top asymmetries analyzed in Section \ref{sec:FBTevatron}. In the last column we report the correspondent SM predictions.}}\label{tab:TevatronData}
\end{table}
In particular we simulate both processes in Eq. (\ref{eq:TevatronProcessA}) and (\ref{eq:TevatronProcessB}) using MadGraph\,5 \cite{Alwall:2011uj} at $\sqrt{s}=1.96$ TeV with fixed factorization and renormalization scales $\mu_{F}=\mu_{R}=m_{t}$. We take for the top mass $m_{t}=172.5$ GeV, and we use CTEQ6.6M Parton Distribution Functions (PDF's) \cite{Kretzer:2003it,CTEQ}  with $\alpha_s(m_Z)=0.118$.
On the other hand, with the help of MSTW~\cite{Martin:2009iq}, we convolute with the parton distribution function the partial differential cross-section
for $u(p_1)\bar{u}(p_2)\rightarrow t(k_1)\bar{t}(k_2)$ from Ref.~\cite{Shu:2009xf},
\be
\label{eq:tt_X-section}
\frac{d\sigma}{d\cos\theta} =\frac{\beta}{1152\pi s} \left[ 8 g_S^4 ( 1+c_\theta^2 +4 m^2)+ 16 \lambda^2  g_S^2 s \frac{(1-c_\theta)^2+4m^2}{ t_{\phi}}
+ 36\lambda^4 \frac{ s^2(1-c_\theta)^2}{t_{\phi}^2}\right],
\ee
where $s \equiv (p_1+p_2)^2$, $t \equiv (p_1-k_1)^2$, $t_t \equiv t-m_t^2$, $m \equiv m_t/\sqrt{s}$, $t_\phi \equiv t-m_0^2$, and  $g_S$ is the QCD coupling constant. $\theta$ and $\beta$ are the scattering angle between the outgoing top and incoming quark and top quark velocity, defined in the partonic center-of-mass frame, and $c_\theta=\beta\cos\theta$. Note that the second term arises from the mixing between the gluon- and $\phi$-exchange and the third term from $\phi$-exchange only, that are terms responsible for the top-quark FB asymmetries.
$A_{\rm FB}^{t\overline{t}}$ can be obtained by integrating over the forward ($\cos\theta >0$) or backward ($\cos\theta < 0$) region. The resulting values for $A_{\rm FB}^{t\overline{t}}$ are in agreement with those from MadGraph\,5.

\subsubsection{Other top-related constraints}
\label{subsec:OtherConstraints}
\begin{enumerate}
\item {\underline{The $t\overline{t}$ total production cross section}}.
The most puzzling feature of the top FB asymmetry is that the new physics responsible for it must not significantly alter the SM prediction for the $t\overline{t}$ total inclusive production cross section at NLO QCD $\sigma_{\rm SM, NLO}^{t\overline{t},\,{\rm Tev}}=6.9$ pb \cite{xSectionOnLine}. The latter, in fact, is in good agreement with the value measured by CDF \cite{CDFtotal},
\begin{equation}\label{eq:CDFtotoal}
\sigma_{\rm CDF}^{t\overline{t}}=7.5\pm 0.48~{\rm pb}~.
\end{equation}
In order to keep under control this constraint first we compute the LO SM total production cross section $\sigma_{\rm SM, LO}^{t\overline{t},\,{\rm Tev}}=5.68$ pb, and then, following \cite{Duffty:2012zz}, we compare Eq. (\ref{eq:CDFtotoal}) with
\begin{equation}\label{eq:NLONPTevatron}
\sigma^{t\overline{t},\,{\rm Tev}}_{\rm NP, NLO}\equiv \sigma_{\rm NP, LO}^{t\overline{t},\,{\rm Tev}}-\sigma_{\rm SM, LO}^{t\overline{t},\,{\rm Tev}}+
\sigma_{\rm SM, NLO}^{t\overline{t},\,{\rm Tev}}~,
\end{equation}
where $\sigma_{\rm NP, LO}^{t\overline{t},\,{\rm Tev}}$ is the total $t\overline{t}$ cross section computed by MadGraph\,5 at the tree level including the New Physics (NP)  contributions related to the t-channel exchange of the scalar doublet $\phi$. In this way we hope to extract at least a universal part of the NLO QCD corrections.
\item {\underline{The top decay width}}. If the neutral component of the scalar doublet in Eqs. (\ref{eq:ScalarDoublet},\ref{eq:PhenoLagrangian}) is light enough, then the decay channel $t\to u\phi^0$ is kinematically allowed, particularly from \cite{Blum:2011fa}
\begin{equation}
\Gamma_{\rm NP}^{t}\equiv\Gamma(t\to u\phi^0)=\frac{m_t |\lambda|^2|V_{tb}|^2}{8\pi}\left(1-\frac{m_0^2}{m_{t}^2}\right)^2~.
\end{equation}
The SM prediction for the decay width of the top quark, including NLO QCD corrections, is $\Gamma_{\rm SM}^t\approx 1.3$ GeV \cite{Jezabek:1988iv}, while the measured value is $\Gamma_{{\rm D}\emptyset}^t=1.99^{+0.69}_{-0.55}$ GeV \cite{Abazov:2010tm}. We show the contour of $\Gamma_{\rm SM}^t+\Gamma_{\rm NP}^{t}$ reproducing the experimental width $\Gamma_{{\rm D}\emptyset}^t$  in the left panel of Fig. \ref{fig:TopDecay}. It turns out that, in a large slice of the parameter space $(m_0,|\lambda|)$ of interest, the existence of the scalar doublet can mitigate the tension between the SM prediction and the observed value. It is important to emphasize that the experimental researches on the top quark are mainly based on $Wb$ final states. In this paper, we simply assume that events from $t\to u\phi^0$ will not pass event selections optimized from $Wb$ final states. Therefore, for $m_0+ m_u < m_t $, the resulting $\sigma^{t\overline{t}}$ and $\Gamma^{t}$ will be re-scaled by $(\Gamma_{\rm SM}^{t}/\Gamma_{\rm SM}^{t}+\Gamma_{\rm NP}^{t})^2$ and $\Gamma_{\rm SM}^{t}/\Gamma_{\rm SM}^{t}+\Gamma_{\rm NP}^{t}$, respectively. In this case we need the larger $\lambda$ to account for the reduction to satisfy the experimental data.

\end{enumerate}

\subsubsection{Results}\label{subsec:TevatronResults}
\begin{figure}[!htb!]
  \begin{minipage}{0.5\textwidth}
   \centering
   \includegraphics[scale=0.6]{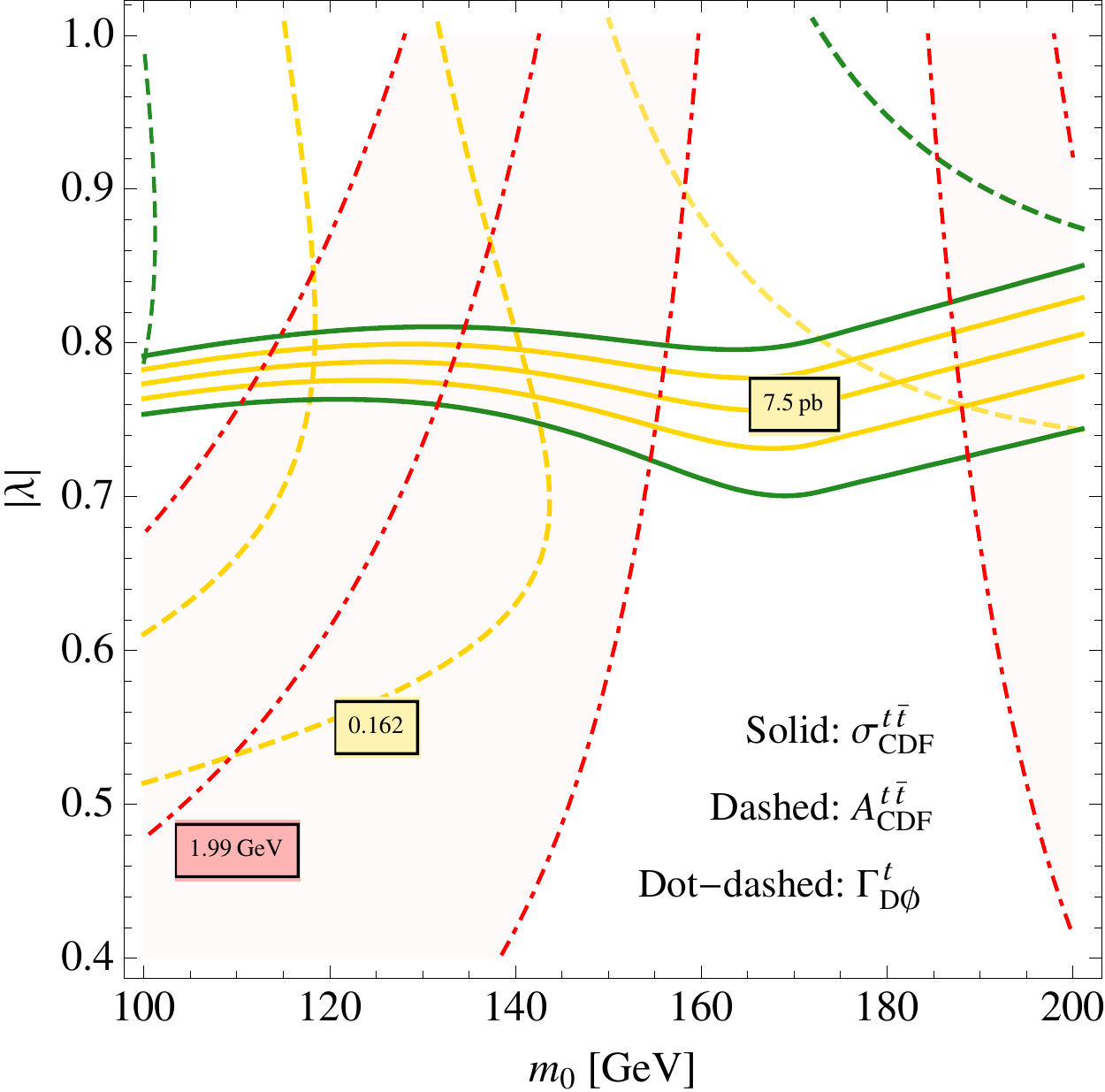}
    \end{minipage}\hspace{1 cm}
   \begin{minipage}{0.25\textwidth}
    \centering
    \includegraphics[scale=0.65]{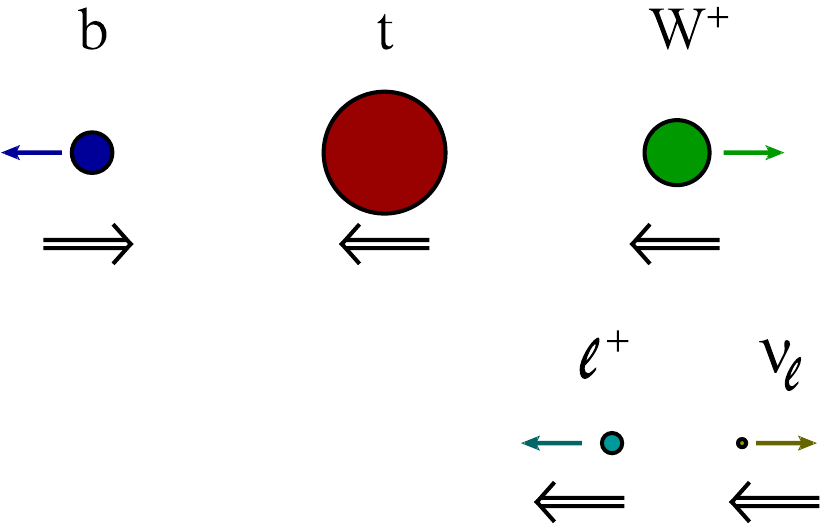}
    \end{minipage}
 \caption{\emph{{\underline{Left panel}}:
 contour plot in the  $(m_0,|\lambda|)$ plane; we show i) the contours in which $\Gamma^t_{\rm SM}+\Gamma^t_{\rm NP}$ reproduces the measured value $\Gamma_{{\rm D}\emptyset}^t=1.99^{+0.69}_{-0.55}$ GeV (red dot-dashed lines) ii) the CDF values, together with $1\,\sigma$ and $2\,\sigma$ bands, for $A_{\rm FB}^{t\overline{t}}$ (dashed lines) and $\sigma^{t\overline{t}}$ (solid lines). Only central values are labelled. {\underline{Right panel}}: sketch of the leptonic top decay; double arrows represent helicity.}}
 \label{fig:TopDecay}
\end{figure}

We here present our results for the scalar doublet $\phi$ in terms of the FB asymmetries. We start considering the inclusive FB asymmetry $A_{\rm FB}^{t\overline{t}}$; according to Tab. \ref{tab:TevatronData}, in fact, this is the only result confirmed by both CDF and ${\rm D}\emptyset$ collaborations. In order to compare the theoretical prediction with the correspondent experimental result, we define
\begin{equation}\label{eq:AsymmetryCombined}
A_{\rm FB}^{t\overline{t}}\equiv \frac{A_{\rm FB,NP}^{t\overline{t}}\times\sigma_{\rm NP, LO}^{t\overline{t},\,{\rm Tev}}
+A_{\rm FB,SM}^{t\overline{t}}\times\sigma_{\rm SM, NLO}^{t\overline{t},\,{\rm Tev}}
}{\sigma^{t\overline{t},\,{\rm Tev}}_{\rm NP, NLO}}~,
\end{equation}
where $A_{\rm FB,NP}^{t\overline{t}}$ is the asymmetry generated by the NP interactions, and $A_{\rm FB,SM}^{t\overline{t}}=0.088$ is the SM contribution. We compute $A_{\rm FB,NP}^{t\overline{t}}$  using the events generated by MadGraph\,5, as previously discussed.
\begin{figure}[!htb!]
  \begin{minipage}{0.4\textwidth}
   \centering
   \includegraphics[scale=0.65]{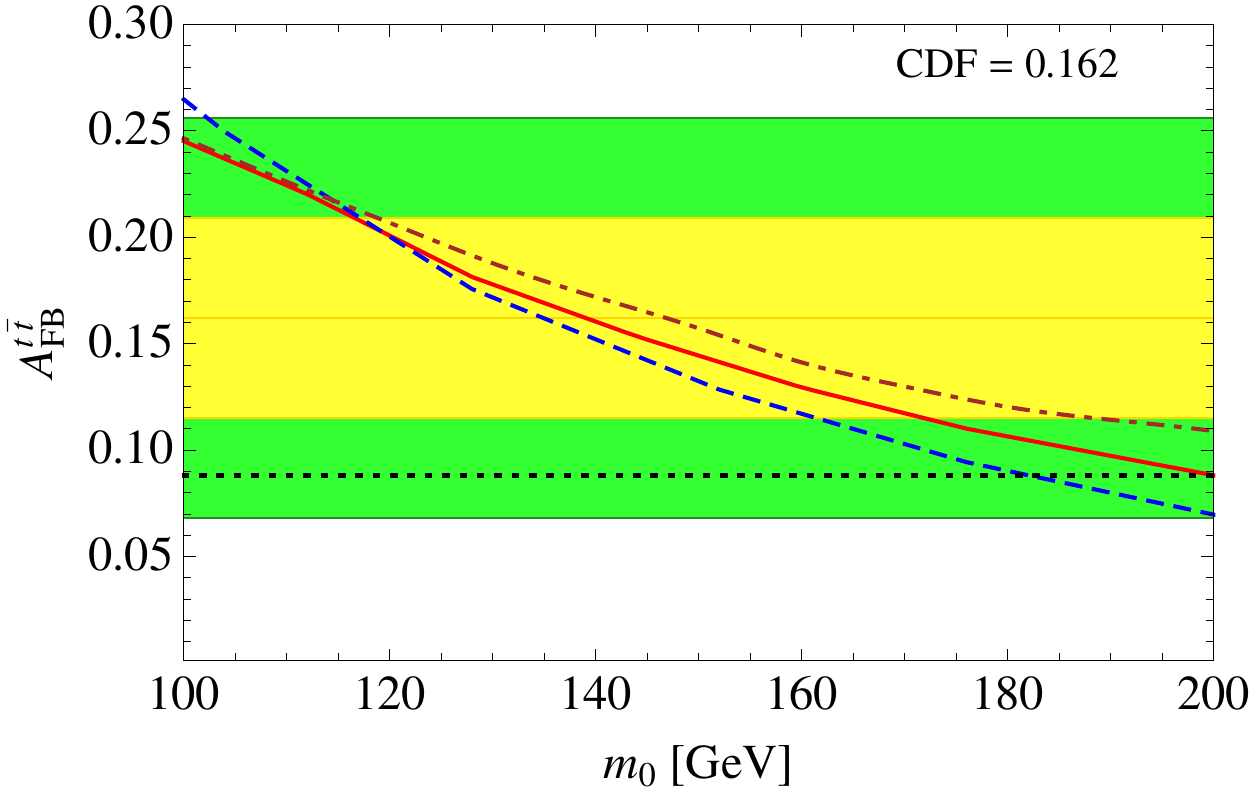}
    \end{minipage}\hspace{2 cm}
   \begin{minipage}{0.4\textwidth}
    \centering
    \includegraphics[scale=0.65]{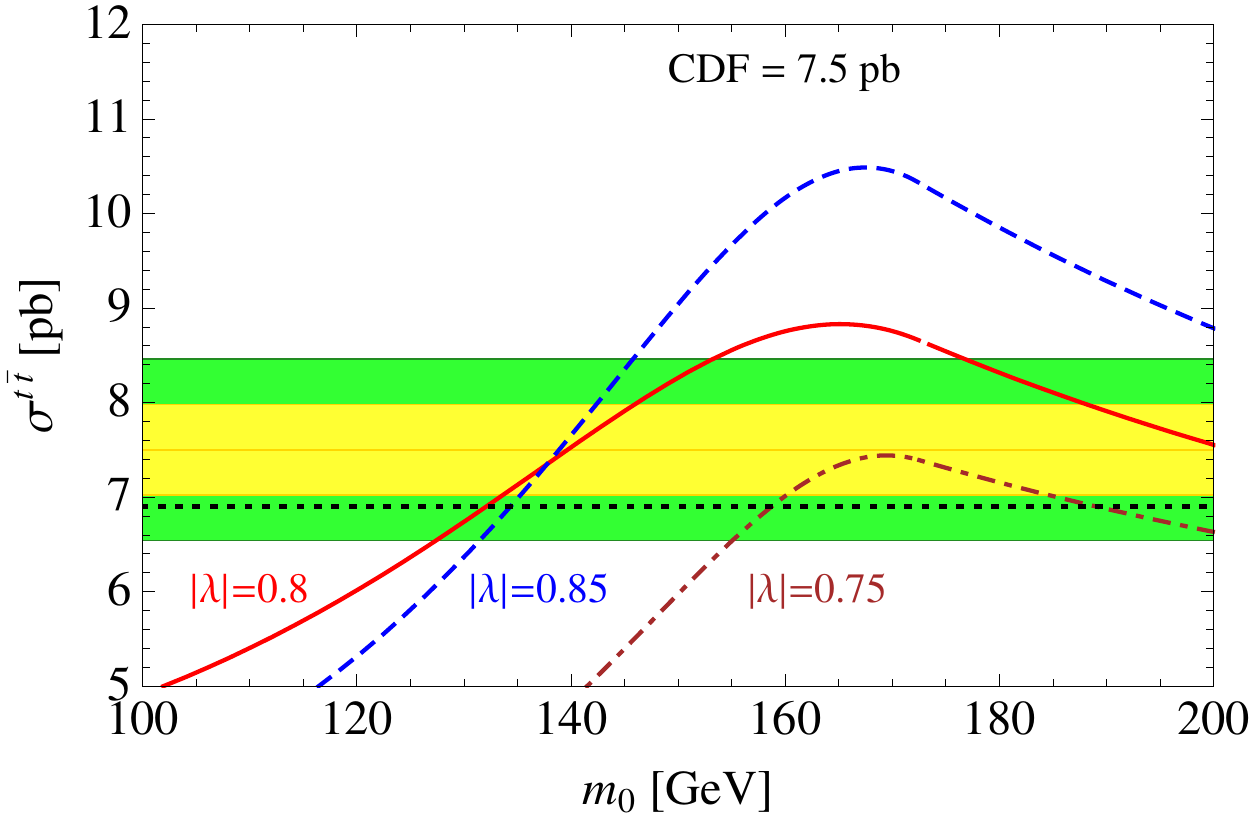}
    \end{minipage}
 \caption{\emph{{\underline{Left panel}}: inclusive FB asymmetry $A_{\rm FB}^{t\overline{t}}$.  {\underline{Right panel}}: $t\overline{t}$ total production cross section. We compare the latest CDF results with the prediction obtained adding to the SM asymmetry (black dotted line) the contribution from the scalar doublet $\phi$ [see Eq. (\ref{eq:NLONPTevatron}) for the $t\overline{t}$ total production cross section and Eq. (\ref{eq:AsymmetryCombined}) for the FB asymmetry]. We analyze three different values for the coupling $\lambda$ in Eq. (\ref{eq:PhenoLagrangian}): $|\lambda|=0.8$ (solid red line),  $|\lambda|=0.85$ (dashed blue line), and  $|\lambda|=0.75$ (dot-dashed brown line). Also shown are the $1\,\sigma$ (yellow) and $2\,\sigma$ (green) regions from the experimental measurements.}}
 \label{fig:TevatronFB}
\end{figure}
 Our results on the asymmetries are shown in the left panel of Fig. \ref{fig:TevatronFB}, together with the latest CDF measurement \cite{Aaltonen:Note}. We analyze three benchmark values for the NP coupling, namely $|\lambda|=0.7,0.75,0.8$, and in the right panel of the same figure we show the corresponding $t\overline{t}$ total production cross sections according to Eq. (\ref{eq:NLONPTevatron}). The impact of the scalar doublet on this last observable is particularly relevant, due to the small values of $m_0$ under scrutiny, thus resulting in large deviations from the experimental cross section.
By contrast, the desired FB asymmetry can be easily reproduced. It turns out that, for instance, within $1\,\sigma$ deviation it is possible to obtain the correct value for $A_{\rm FB}^{t\overline{t}}$ in the range $m_{0}\approx [130,190]$ GeV considering the small window of coupling $\lambda \approx [0.75,0.85]$. Notice that, taking into account the top decay width $\Gamma^{t}_{\rm SM}+\Gamma^{t}_{\rm NP}$ in the same range of couplings, a mass $m_{0}\lesssim 160$ GeV is required to reproduce the measured value $\Gamma_{{\rm D}\emptyset}^t$. A broader view on the allowed values is shown in the left panel of Fig. \ref{fig:TopDecay}, where we present the contours in the $(m_0,|\lambda|)$ plane - together with the $1\,\sigma$ and $2\,\sigma$ bands based on the CDF measurements - for $A_{\rm FB}^{t\overline{t}}$ and $\sigma^{t\overline{t}}$.\\
\begin{figure}
  \begin{minipage}{0.4\textwidth}
   \centering
   \includegraphics[scale=0.65]{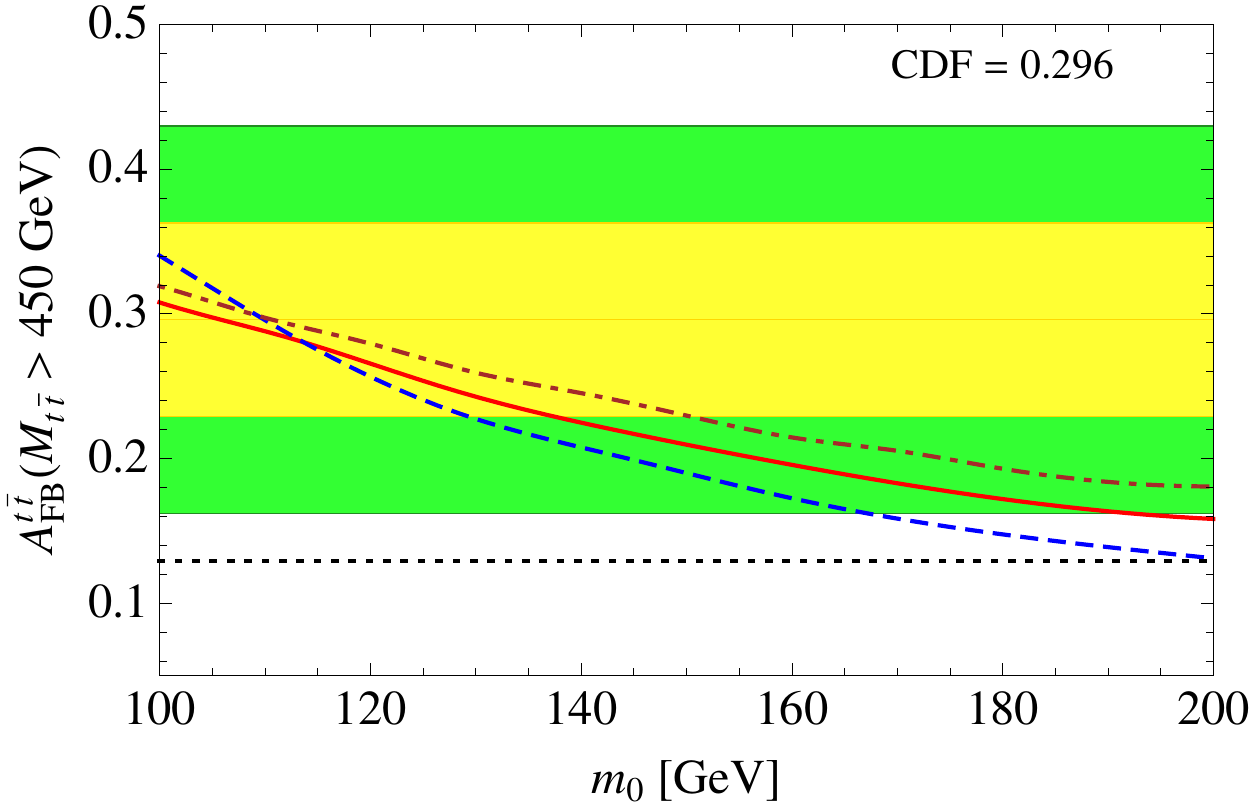}
    \end{minipage}\hspace{2 cm}
  \begin{minipage}{0.4\textwidth}
  \centering
    \includegraphics[scale=0.65]{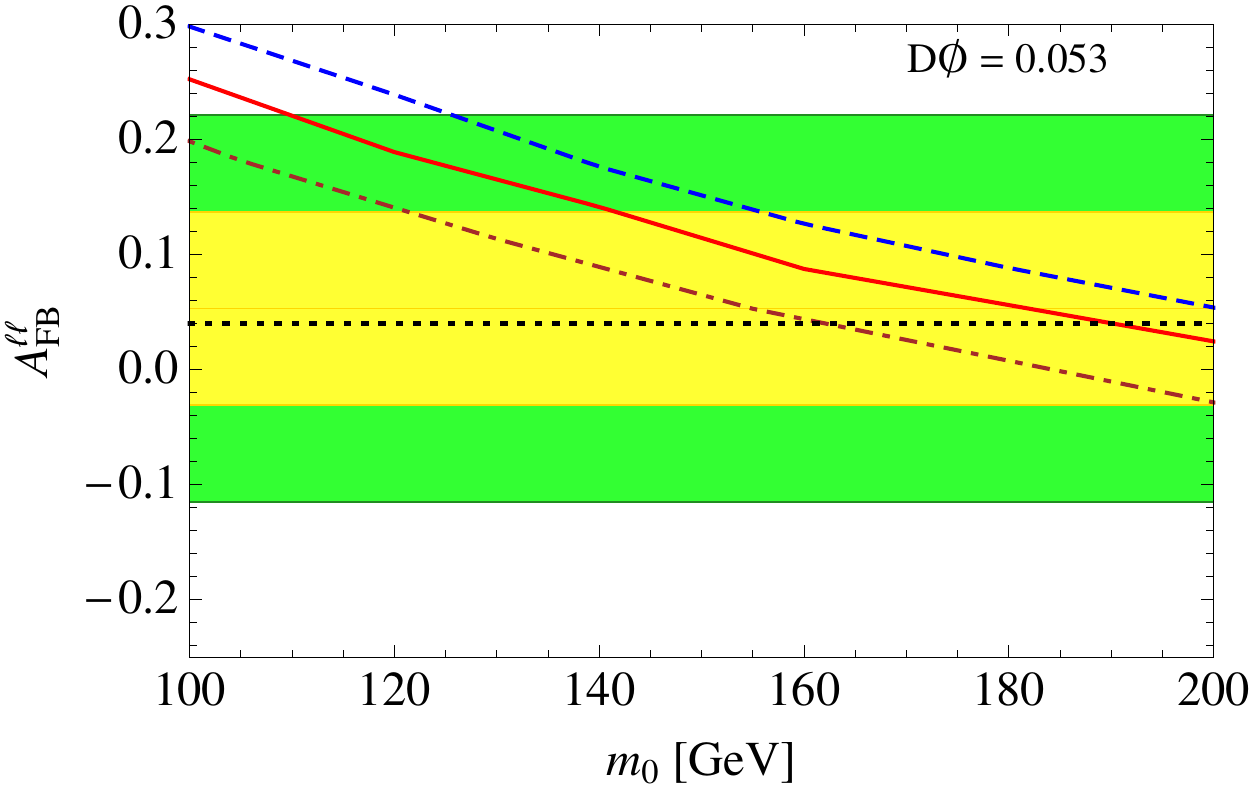}
    \end{minipage}
 \caption{\emph{{\underline{Left panel}}: FB asymmetry  $A_{\rm FB}^{t\overline{t}}(M_{t\overline{t}}> 450\,{\rm GeV})$. We compare the latest CDF result with the prediction obtained adding to the SM asymmetry (black dotted line) the contribution from the scalar doublet $\phi$. Colors and labels follow Fig. \ref{fig:TevatronFB}. {\underline{Right panel}}:  inclusive leptonic FB asymmetry $A_{\rm FB}^{ll}$. We compare the latest ${\rm D}\emptyset$ result with the prediction obtained adding to the SM asymmetry (black dotted line) the contribution from the scalar doublet $\phi$.}}
 \label{fig:TevatronL}
\end{figure}
Furthermore, in terms of the top FB asymmetry for large invariant mass $M_{t\overline{t}}>450$ GeV, the latest result from the CDF collaboration \cite{Aaltonen:Note}, based on a larger data sample, is significantly smaller with respects to the $3.4\,\sigma$ deviation from the SM reported in \cite{Aaltonen:2011kc}. Nevertheless a mild tension ($\sim 2.4\,\sigma$) still persists. We show our results for this observable in the left panel of Fig. \ref{fig:TevatronL}, where we add the contribution of the scalar doublet $\phi$ to the SM asymmetry. With $\lambda\in [0.75,0.85]$ and $m_0\lesssim 150$ GeV, the scalar doublet $\phi$ can account for the data within $1\,\sigma$ error.\\
Finally, we would like to comment about the leptonic FB asymmetry. In order to combine the SM prediction with the NP contribution first we introduce the $K$-factor $K_{\rm Tev}\equiv\sigma_{\rm SM, NLO}^{t\overline{t},\,{\rm Tev}}/\sigma_{\rm SM, LO}^{t\overline{t},\,{\rm Tev}}\approx 1.214$, then we define, in close analogy with Eq. (\ref{eq:AsymmetryCombined})
\begin{equation}\label{eq:TevatronLeptonicAsymmetry}
A_{\rm FB}^{ll}\equiv \frac{A_{\rm FB,NP}^{ll}\times
\sigma_{\rm NP, LO}^{ll,\,{\rm Tev}}+A_{\rm FB,SM}^{ll}\times
K_{\rm Tev}\times\sigma_{\rm SM, LO}^{ll,\,{\rm Tev}}
}{\sigma_{\rm NP, NLO}^{ll,\,{\rm Tev}}}~,
\end{equation}
where $\sigma_{\rm NP, NLO}^{ll,\,{\rm Tev}}\equiv \sigma_{\rm NP, LO}^{ll,\,{\rm Tev}}-\sigma_{\rm SM, LO}^{ll,\,{\rm Tev}}(1-K_{\rm Tev})$.
In Eq. (\ref{eq:TevatronLeptonicAsymmetry}) $A_{\rm FB,NP}^{ll}$ is the leptonic FB asymmetry generated by the exchange of the scalar doublet $\phi$ while  $A_{\rm FB,SM}^{ll}=0.048$ is the SM contribution.  $\sigma_{\rm NP, LO}^{ll,\,{\rm Tev}}$ and $\sigma_{\rm SM, LO}^{ll,\,{\rm Tev}}$ refer, respectively, to the tree level cross sections for the di-leptonic process in Eq. (\ref{eq:TevatronProcessB}) with and without the NP contribution. $A_{\rm FB,NP}^{ll}$, $\sigma_{\rm NP, LO}^{ll,\,{\rm Tev}}$, and $\sigma_{\rm SM, LO}^{ll,\,{\rm Tev}}$ are obtained from MadGraph\,5, and in particular we impose the following cuts on the final state
\begin{equation}
p_T^l\geqslant 20~{\rm GeV}~,~~~|\eta_l|\leqslant 2.0~,~~~p_T^b\geqslant 20~{\rm GeV}~,~~~|\eta_b|\leqslant 2.0~,~~~\slashed{E}_T\geqslant 25~{\rm GeV}~,
\end{equation}
where $p_T^j$ and $\eta_j$ refer, respectively, to the transverse momentum and pseudo-rapidity of the particle $j$. We show our result in the right panel of Fig. \ref{fig:TevatronL} with the latest ${\rm D}\emptyset$ result \cite{Abazov:2011rq}. The inclusion of the scalar doublet can significantly modify the SM prediction but remain consistent with the experimental result.\\
To explain in a naive way the correlation between $A_{\rm FB}^{t\overline{t}}$ in Fig. \ref{fig:TevatronFB}  and $A_{\rm FB}^{\ell\ell}$ in Fig. \ref{fig:TevatronL}, we start
considering the top-quark decay at rest as shown in the right panel of Fig. \ref{fig:TopDecay}, where double arrows represent helicity (also chirality for massless particles).
In the limit of $m_t \sim m_{W^+} \gg m_b$, $m_{\ell^+}$ and $m_{\nu}$ and due to the fact the weak interactions involve left-handed chirality only,  it is clear that to conserve the angular momentum, the charged lepton tends to move along the direction of the top-quark spin. Therefore, for the right- (left-) handed helicity top quark, i.e., to boost the top quark along (against) the spin direction, the resulting charged lepton moves along (against) the direction of the top quark. For the highly boosted top-quark, however, the lepton is prone to moves along the top quark direction regardless of the helicity.\footnote{Please see, for example, Ref.~\cite{Berger:2012tj} for more detailed discussion.} Moreover, in Eq. (\ref{eq:tt_X-section}), the second term on the right hand side becomes dominant for $\lambda \ll 1$ and the third term becomes comparable when $\lambda \gtrsim 1$. At the same time the second term is roughly proportional to the partonic center-of-mass energy squared, $s$, and the third term proportional to $s^2$. In other words, for $\lambda \ll 1$, there are more low energy top-quarks than the case of $\lambda \gtrsim 1$, which in turn implies, the more resulting charged lepton ($\ell^+$) move against top-quarks in the $\lambda \ll 1$ case. Therefore, for $\lambda=0.75$ and $m_0=100$ GeV, $A_{\rm FB}^{\ell\ell}$ is negative and $A_{\rm FB}^{t\overline{t}}$ is positive while for $\lambda=0.85$ they are both positive.
As $m_0$ increases, $A_{\rm FB}^{\ell\ell}$ decreases due to decreased $A_{\rm FB}^{t\overline{t}}$.\\
Let us now conclude this Section with a contour plot in the $(m_0,|\lambda|)$ plane, shown in the left panel of Fig. \ref{fig:TopDecay}. The combined analysis involving the inclusive FB asymmetry, the $t\overline{t}$ total production cross section and the top decay width selects a small slice of the considered parameter space $(m_0,|\lambda|)$.
In Section \ref{sec:SingleTopProduction} we will study the role of the scalar doublet $\phi$ in the single top production at LHC, with the benchmark point $m_0=140$ GeV, $|\lambda|=0.8$.

\subsection{The charge and leptonic forward-backward asymmetries at the LHC}
\label{sec:FBLHC}
At the LHC, the FB asymmetry defined in Eq. (\ref{eq:FBdefinition}) vanishes because the proton-proton initial state is symmetric. Due to the composition of proton, however, in terms of quarks, QCD predicts a charge asymmetry for the $t\overline{t}$ pair production  generated from $q\overline{q}$ and $gq$ ($g\overline{q}$):
antitop quarks are produced more likely in the central region, with top quarks more abundant at large positive and negative rapidities \cite{Kuhn}. On the other hand this effect is drastically reduced because
top quark pair production at the LHC is dominated by gluon-gluon fusion ($84\%$ at $\sqrt{s}=10$ TeV and $90\%$ at $\sqrt{s}=14$ TeV), a charge symmetric process.
The exchange of the scalar doublet $\phi$ contributes at the tree level to the process $u\overline{u}\to t\overline{t}$, producing a non-vanishing contribution to the charge asymmetry which is, however, diluted by the dominance of gluon-gluon fusion. Nevertheless we will still evaluate this charge asymmetry in light of the latest experimental results from ATLAS and CMS collaborations. After a more detailed introduction in Section \ref{subsec:LHCAsymmetry}, we show our results in Section \ref{subsec:LHCResults}.

\subsubsection{Description of the asymmetries}
\label{subsec:LHCAsymmetry}
For top-quark pair production at the LHC
\begin{equation}\label{eq:LHCProcessA}
p\,p\to t\,\overline{t},
\end{equation}
the charge asymmetry in the laboratory frame is defined as
\begin{equation}
A_{\rm C}^{t\overline{t}}\equiv \frac{N(\Delta|y|>0)-N(\Delta|y|<0)}{
N(\Delta|y|>0)+N(\Delta|y|<0)}~,
\end{equation}
where $\Delta|y|\equiv |y_t|-|y_{\overline{t}}|$.\footnote{Notice that at the LHC an asymmetry based on the variable $\Delta y=y_t-y_{\overline{t}}$, used at the Tevatron in the context of the FB asymmetry, would vanish due to the symmetry of the incoming beams.} On the other hand, considering
a di-leptonic final state from top decay
\begin{equation}\label{eq:LHCProcessB}
p\,p\to t\,\overline{t}\to l^{+}\,l^{-}\,j_{b}\,j_{\overline{b}},
\end{equation}
a leptonic asymmetry in the laboratory frame is defined as
\begin{equation}
A_{\rm C}^{ll}\equiv \frac{N(\Delta |\eta_l|>0)-N(\Delta |\eta_l|<0)}{N(\Delta |\eta_l|>0)+N(\Delta |\eta_l|<0)}~,
\end{equation}
where $\Delta|\eta_l|\equiv |\eta_{l^+}|-|\eta_{l^{-}}|$.
We summarize in Tab. \ref{tab:LHCAsymmetry} the latest results from ATLAS and CMS experiments together with the correspondent SM predictions.
\begin{table}[!htb!]
\begin{center}
\begin{tabular}{ | c | c | c | c |}
\hline
$\sqrt{s}=7$ TeV  & ATLAS \cite{ChargeAsymmetryATLAS} & CMS \cite{ChargeAsymmetryCMS} & SM (MC@NLO) \cite{ChargeAsymmetryATLAS}  \\ \hline
$A_{\rm C}^{t\overline{t}}$  & $0.057\pm0.024\pm0.015$ & $0.004\pm0.010\pm0.001$ & $0.006\pm 0.002$  \\ \hline
 $A_{\rm C}^{ll}$ & $0.023\pm0.012\pm0.008$ & $\ast$ & $0.004\pm0.001$  \\ \hline
 \end{tabular}
\end{center}
\caption{\emph{Unfolded experimental results from ATLAS and CMS collaborations for the top asymmetries analyzed in Section \ref{sec:FBLHC}. In the last column we report the correspondent SM predictions.}}\label{tab:LHCAsymmetry}
\end{table}
Notice that considering the $t\overline{t}$ total production cross section the results from ATLAS \cite{Calkins:2011mx} and CMS \cite{Chatrchyan:2011nb} at $\sqrt{s}=7$ TeV are $ \sigma^{t\overline{t}}_{\rm ATLAS}=176 \pm 5^{+13}_{-10}\pm 7$ pb, and  $ \sigma^{t\overline{t}}_{\rm CMS}=168 \pm 18\pm 14\pm 7$ pb while for the NLO QCD SM from \cite{Langenfeld:2009wd} we have $\sigma_{\rm SM,NLO}^{t\overline{t},\,{\rm LHC}}=165.80^{+4.44}_{-6.99}\pm 9.10\pm 11.6$ pb.
Like the Tevatron asymmetries, we simulate both processes in Eq. (\ref{eq:LHCProcessA}) and (\ref{eq:LHCProcessB}) using MadGraph\,5 \cite{Alwall:2011uj} at $\sqrt{s}=7$ TeV with fixed factorization and renormalization scales $\mu_{F}=\mu_{R}=m_{t}$. We take for the top mass $m_{t}=172.5$ GeV, and we use CTEQ6.6M Parton Distribution Functions (PDF's) \cite{Kretzer:2003it,CTEQ}  with $\alpha_s(m_Z)=0.118$.

\subsubsection{Results}
\label{subsec:LHCResults}
\begin{figure}[!htb!]
  \begin{minipage}{0.4\textwidth}
   \centering
   \includegraphics[scale=0.65]{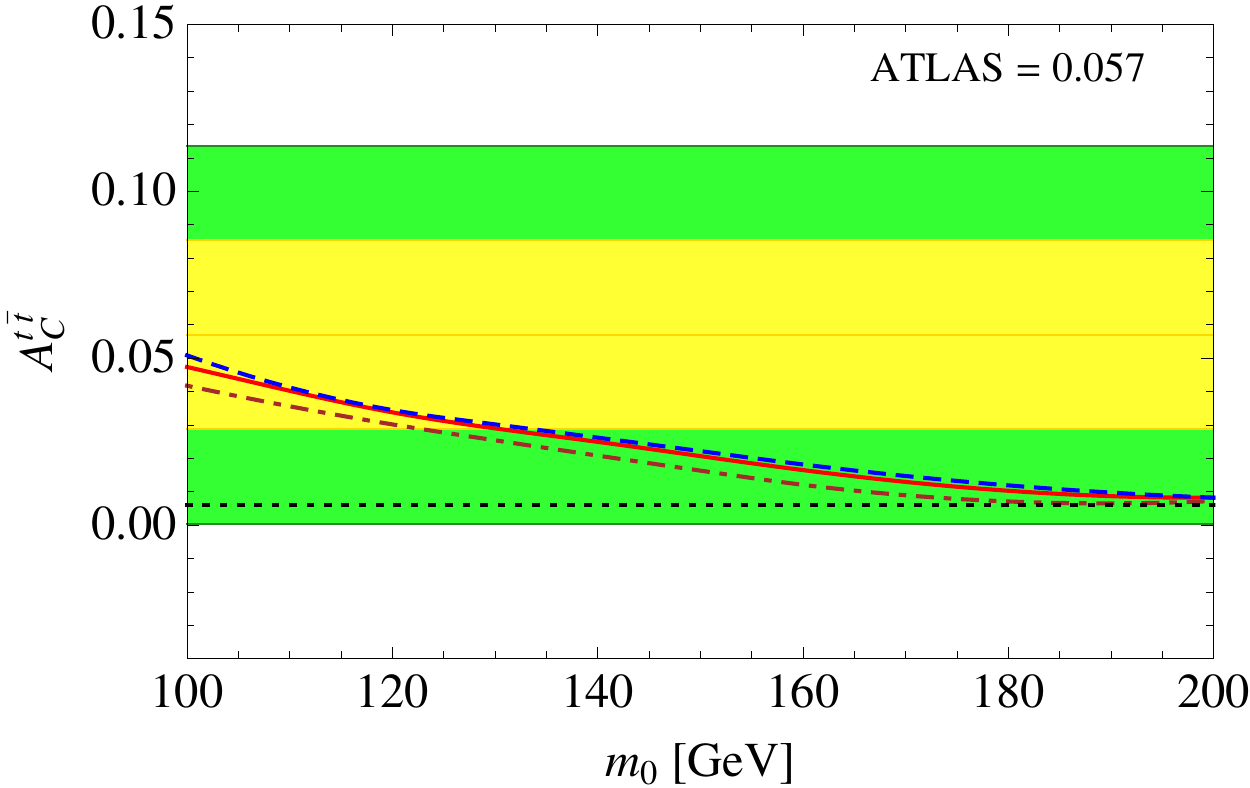}
    \end{minipage}\hspace{2 cm}
   \begin{minipage}{0.4\textwidth}
    \centering
    \includegraphics[scale=0.65]{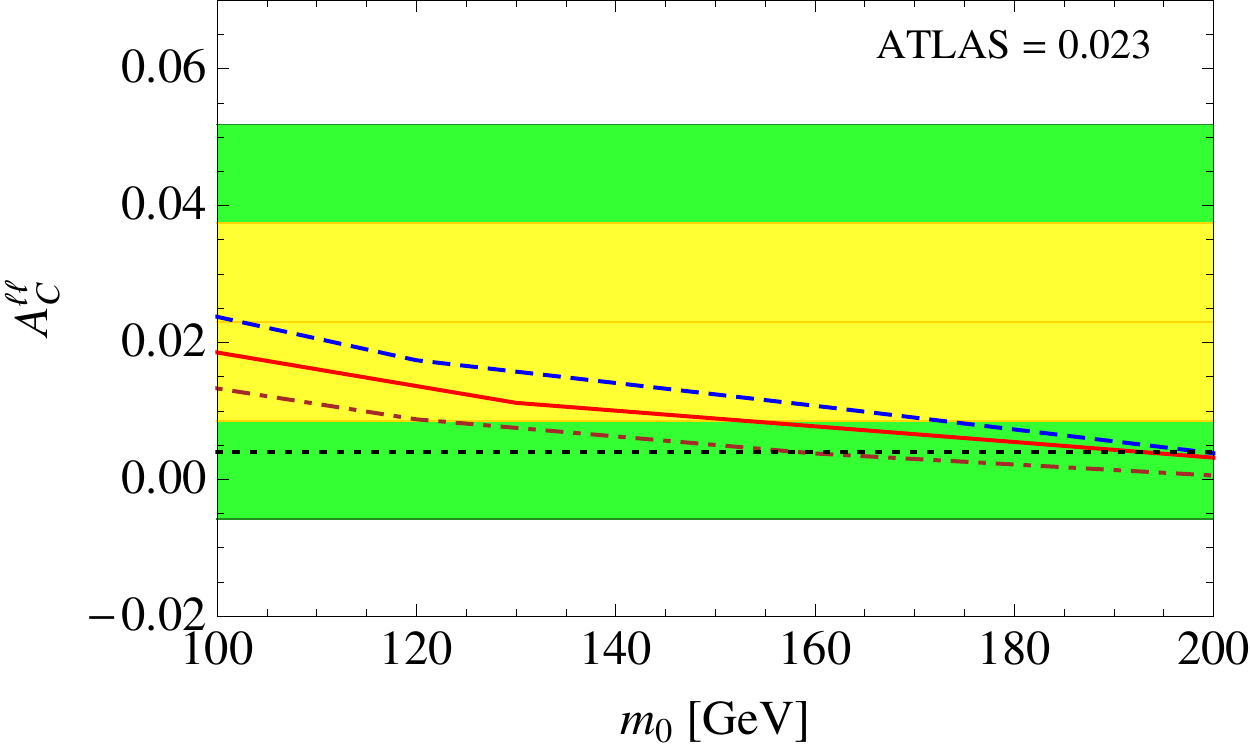}
    \end{minipage}
 \caption{\emph{Left panel: Right panel}}
 \label{fig:LHCCharge}
\end{figure}

Let us start our analysis considering the top charge asymmetry. Following the same strategy implemented in Section \ref{sec:FBTevatron}, we define
\begin{equation}\label{eq:ChargeAsymmetryCombined}
A_{\rm C}^{t\overline{t}}\equiv \frac{A_{\rm C,NP}^{t\overline{t}}\times\sigma_{\rm NP, LO}^{t\overline{t},\,{\rm LHC}}
+A_{\rm C,SM}^{t\overline{t}}\times\sigma_{\rm SM, NLO}^{t\overline{t},\,{\rm LHC}}
}{\sigma^{t\overline{t},\,{\rm LHC}}_{\rm NP, NLO}}~,
\end{equation}
where [see Eq. (\ref{eq:NLONPTevatron})]
\begin{equation}\label{eq:Charge2}
\sigma^{t\overline{t},\,{\rm LHC}}_{\rm NP, NLO}\equiv \sigma_{\rm NP, LO}^{t\overline{t},\,{\rm LHC}}-\sigma_{\rm SM, LO}^{t\overline{t},\,{\rm LHC}}+
\sigma_{\rm SM, NLO}^{t\overline{t},\,{\rm LHC}}~.
\end{equation}
 We show our result for the charge asymmetry in the left panel of Fig. \ref{fig:LHCCharge}. We find that the presence of the extra scalar doublet $\phi$ can slightly increase by few percent the value of the SM charge asymmetry pointing towards the central value measured by ATLAS. This effect starts to be significant for $|\lambda|\gtrsim 0.8$ and $m_{0}\lesssim 140$ GeV. Due to the dominance of gluon-gluon fusion on the total production cross section, this range of values is consistent with $ \sigma^{t\overline{t}}_{\rm ATLAS/CMS}$; for instance, taking $|\lambda|=0.8$, $m_0=100$ GeV  and using Eq. (\ref{eq:Charge2}) we find $\sigma^{t\overline{t},\,{\rm LHC}}_{\rm NP, NLO}\simeq 174.9$ pb.
For the leptonic charge asymmetry, with $K_{\rm LHC}\equiv\sigma_{\rm SM, NLO}^{t\overline{t},\,{\rm LHC}}/
\sigma_{\rm SM, LO}^{t\overline{t},\,{\rm LHC}}\approx 1.607$, we combine the NP and the SM contribution, defined as
\begin{equation}\label{eq:LHCLeptonicAsymmetry}
A_{\rm C}^{ll}\equiv \frac{A_{\rm C,NP}^{ll}\times
\sigma_{\rm NP, LO}^{ll,\,{\rm LHC}}+A_{\rm C,SM}^{ll}\times
K_{\rm LHC}\times\sigma_{\rm SM, LO}^{ll,\,{\rm LHC}}
}{\sigma_{\rm NP, NLO}^{ll,\,{\rm LHC}}}.
\end{equation}
Besides, MadGraph\,5 is employed to obtain $A_{\rm C,NP}^{ll}$, $\sigma_{\rm NP, LO}^{ll,\,{\rm LHC}}$, and $\sigma_{\rm SM, LO}^{ll,\,{\rm LHC}}$ with the following cuts on the final state,
\begin{equation}
p_T^l\geqslant 20~{\rm GeV}~,~~~|\eta_l|\leqslant 2.5~,~~~p_T^b\geqslant 25~{\rm GeV}~,~~~|\eta_b|\leqslant 2.4~,~~~\slashed{E}_T\geqslant 60~{\rm GeV}~.
\end{equation}
We show our results in the right panel of Fig. \ref{fig:LHCCharge}. The top charge asymmetry generated by the NP contribution is transferred to the di-leptonic final state. The deviation from the SM prediction, however, is really small even with small values of $m_0$ and large couplings $|\lambda|$, thus making it challenging to disentangle NP from the SM.


\section{On the single top production at the LHC}
\label{sec:SingleTopProduction}
\begin{figure}[!htb!]
    \centering
   \includegraphics[scale=0.85]{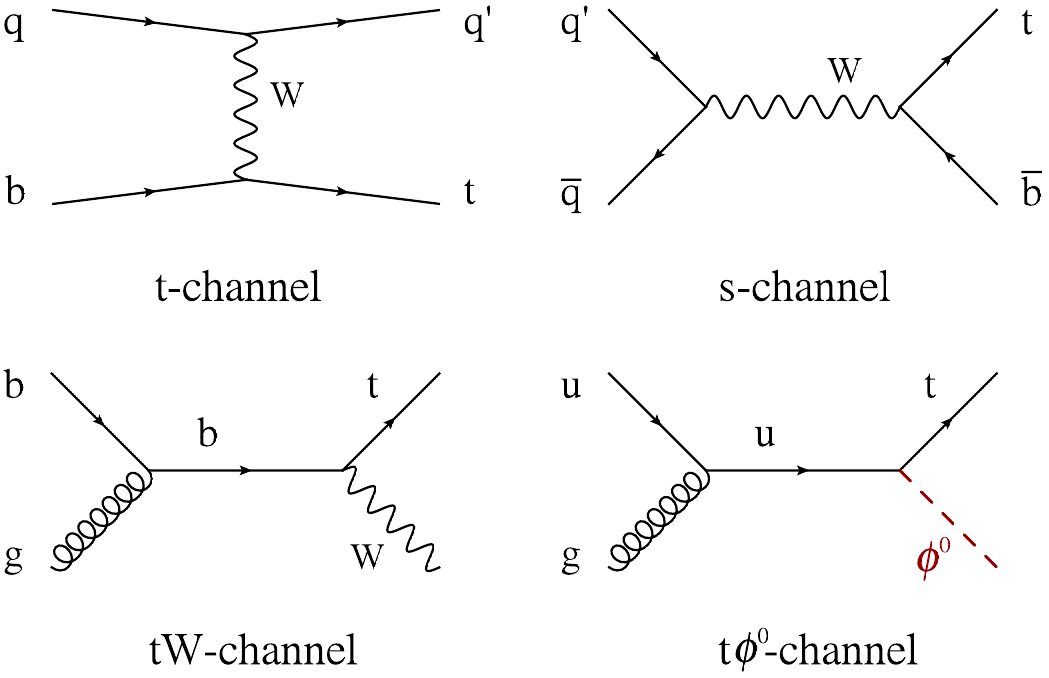}
   \caption{\textit{Representative Feynman diagrams for single top production at the LHC. The first three diagrams describe the usual SM  t-channel, s-channel and tW-channel mechanisms, while the last one corresponds to the associated $\phi^{0}$ production, as explained in the text.
  }}\label{fig:SingleTop}
 \end{figure}

In Section \ref{sec:TopAsymmetries}, we have analyzed the role of the scalar doublet $\phi$ from the point of view of the top asymmetries. Looking at this problem from a broader perspective, however, the need to find other phenomenological fingerprints of its existence arises. Hence, in this Section we shall study the production of single top quarks at the LHC. The single top production at the LHC involves three different sub-processes, namely \textit{t-channel}, \textit{s-channel} and \textit{tW-channel}. Representative Feynman diagrams at parton level for each of these processes are shown in Fig. \ref{fig:SingleTop}.
 Interestingly, the existence of the extra scalar doublet $\phi$ gives rise to a new single top production mechanism mimicking the tW-topology\footnote{tW-channel has a very small cross section at the Tevatron and therefore the event selection for the single top production is not optimized for such a topology. That is the reason why we focus on the LHC.}, as shown in the last diagram in Fig. \ref{fig:SingleTop}. In this case, in fact,  the neutral component of the scalar doublet, instead of a $W$ boson, accompanies the top quark production; in the following we call this new production mechanism \textit{t$\phi^0$-channel}. Moreover, we highlight two important differences.
 \begin{enumerate}[i.]
 \item Due to the peculiar flavor structure of its interactions specified in Eq. (\ref{eq:PhenoLagrangian}), the single top production mechanism in the $t\phi^0$-channel is primed by a up quark instead of a bottom quark for the tW-channel.
 Qualitatively, this simply implies that the parton level cross sections for these two processes are weighted by different PDF's. The $t\phi^0$-channel, as a consequence, is boosted by the contribution of the valence up-quark of the proton.
 \begin{table}
\begin{center}
\begin{tabular}{ | c | c | c | }
\hline
  & $bg\to tW^{-}$& $ug\to t\phi^0$  \\
  \hline
 Parton Level ($\sqrt{\hat{s}}=1$ TeV) & $2.7$ pb & $3.8$ pb  \\
  \hline
PDF's convolution ($\sqrt{s}=7$ TeV)    & $5.2$ pb & $77.8$ pb   \\
 \hline
PDF's convolution ($\sqrt{s}=14$ TeV)    & $30.0$ pb & $286.8$ pb   \\
  \hline
\end{tabular}
\end{center}
\caption{\emph{Cross sections for the scattering processes $bg\to tW^{-}$ (SM tW-channel) and $ug\to t\phi^0$ (t$\phi^0$-channel) at parton level and after PDF's convolution. Analytical results are cross-checked with CalcHEP \cite{Belyaev:2012qa}.}}\label{tab:CrossSectionExample}
\end{table}
We compare in Tab. \ref{tab:CrossSectionExample} the cross sections for the two processes $ug\to t\phi^0$ and $bg\to tW^{-}$ at the parton level and after convolution with the PDF's, taking $m_0=140$ GeV and $|\lambda|=0.8$. We find that they are comparable at the parton level but, as expected, the former is enhanced with respect to the latter after PDF's convolution.

\item  Considering the decay of the neutral component $\phi^0$, only hadronic decays are allowed according to Eq. (\ref{eq:PhenoLagrangian}),
\begin{eqnarray}
\Gamma(\phi^0\to q\overline{u})=\Gamma(\phi^0\to u\overline{q})&=&
\frac{|\lambda|^2|V_{qb}|^2(m_0^2-m_q^2-m_u^2)}{4\pi m_0^3}\times\nonumber\\
&&
\sqrt{\left[m_0^2-(m_q+m_u)^2\right]\left[m_0^2-(m_q-m_u)^2\right]}~,
\end{eqnarray}
with $m_0>m_q+m_u$. Therefore, for $m_0<m_t+m_u$  only decays into light quarks $(u,c)$ are kinematically possible. On the contrary, the associated $W$ in the tW-channel decays into light quarks $(u,d,c,s)$ at the price of a branching ratio $\sim 66\%$.

 \end{enumerate}

\begin{figure}[!htb!]
    \centering
   \includegraphics[scale=0.85]{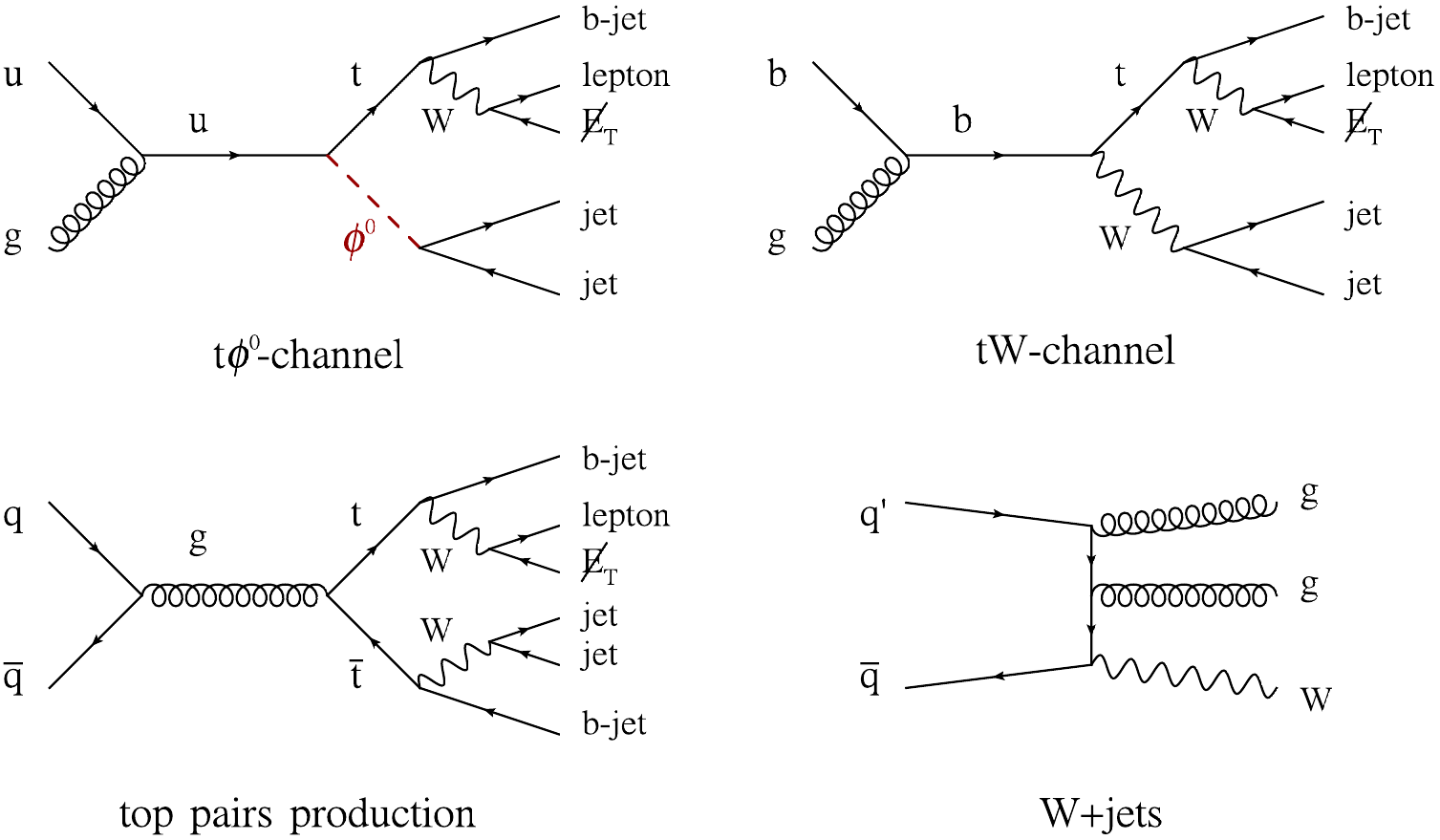}
   \caption{\textit{Representative Feynman diagrams for the semi-leptonic single top production in the t$\phi^0$-channel. Together with the signal we show the irreducible background due to the tW-channel, and the reducible background due to the top pairs production and the $W$+jets final state.
  }}\label{fig:SingleTopProduction}
 \end{figure}

 Bearing in mind these two points, in order to look for an experimental evidence of the t$\phi^0$-channel we focus at the detector level on a semi-leptonic final state. This final state is composed by:
 \begin{itemize}
 \item one lepton (electron or muon), a b-tagged jet, and missing energy from the top decay $t\to bW$ and the subsequent leptonic decay of the $W$ boson;

 \item two jets from the $\phi^0$ decay.

 \end{itemize}
This topology is shown in the upper panel of Fig. \ref{fig:SingleTopProduction} both for the t$\phi^0$-channel and the tW-channel\footnote{ For the diagram 2 and 3 in Fig. \ref{fig:SingleTopProduction}, the interchange of leptonic and hadronic decays of two $W$ bosons have been taken into account in simulation.}. Notice that in this case the tW-channel  plays the role of an irreducible background for the $\phi^0$ signal.\\
Several processes, on the other hand, constitute the reducible background for our analysis. Especially, we take into account the single top production in the t/s-channel, the $W$+jets channel, and the top pairs production. Events from t/s-channel ($W$+jets channel) can have the same final state as the  t$\phi^0$-channel because of  the QCD radiation of an extra jet (b-tagged jet), while for the third case it results from the misidentification of a b quark jet. Representative Feynman diagrams for top pairs production and $W$+jets channels are shown in the lower panel of Fig. \ref{fig:SingleTopProduction}.\\
 Notice that single top production searches in the tW-channel at the LHC have been done both at ATLAS \cite{SingleTopATLAS} and CMS \cite{SingleTopCMS} considering only a di-leptonic final state. These analysis do not apply to our case, and consequently we here try to motivate the necessity to perform detailed experimental studies also considering the single top production in the semi-leptonic tW-channel. Therefore, let us now describe the details of our analysis.\\
Events are generated using MadGraph\,5 \cite{Alwall:2011uj}, and hadronized/showered in Pythia\,6.420 \cite{Sjostrand:2006za}. We use PGS\,4 \cite{PGS4} to perform the fast detector simulation. Considering the $W$+jets channel (t/s-channel) we simulate events with $1,2,3$ jets ($1,2$ jets), using the MLM matching algorithm \cite{Mangano:2006rw}.\\
We select events matching the final state topology of the signal by requiring
\begin{enumerate}
\item exactly one isolated lepton with $p_T>20$ GeV (both for the electron and muon), $|\eta|<2.5$ (muon), $|\eta|<2.47$ (electron);
\item at least three jets with $p_T>30$ GeV and $|\eta|<5$; only one of these jets is b-tagged;
\item $\slashed{E}_T>20$ GeV for the missing energy.

\end{enumerate}
\begin{table}
\begin{center}
\begin{tabular}{ | c | c | c | c | c |c|}
\hline
 $\sqrt{s}=14$ TeV ($7$ TeV) & t$\phi^0$-channel & tW-channel & t/s-channel & $t\overline{t}$  & $W$+jets \\
  \hline
 Cross section [pb] & 150.5 (41.83) & 5.59 (1) & 29.76 (8.47) & 49.03 (9) & 2768 (1166) \\
  \hline
 NOEa/NOEb (\%) & 14.9 (16.19) & 13.9 (13.8) & 2.49 (2.38) & 19 (19.9) & 0.19 (0.13)\\
  \hline
\end{tabular}
\end{center}
\caption{\emph{At 14 (7) TeV, cross sections and cuts efficiency (ratio between the number of events after and before  the cuts) for the signal and background processes.}}\label{tab:Events}
\end{table}
In Tab. \ref{tab:Events} we collect the values of the cross sections and the efficiency of the cuts considering both the signal and the background processes. The efficiency is defined as the ratio between the number of events after cuts (NOEa) with respect to the number of generated events before cuts (NOEb). We show our results considering $\sqrt{s}=7$ TeV with integrated luminosity $L=20$ fb$^{-1}$, and $\sqrt{s}=14$ TeV with $L=100$ fb$^{-1}$. For the signal in the t$\phi^0$-channel we consider a benchmark point with $m_0=140$ GeV and $|\lambda|=0.8$, as suggested from the analysis of the top asymmetries.\\
\begin{figure}
  \begin{minipage}{0.4\textwidth}
   \centering
   \includegraphics[scale=0.65]{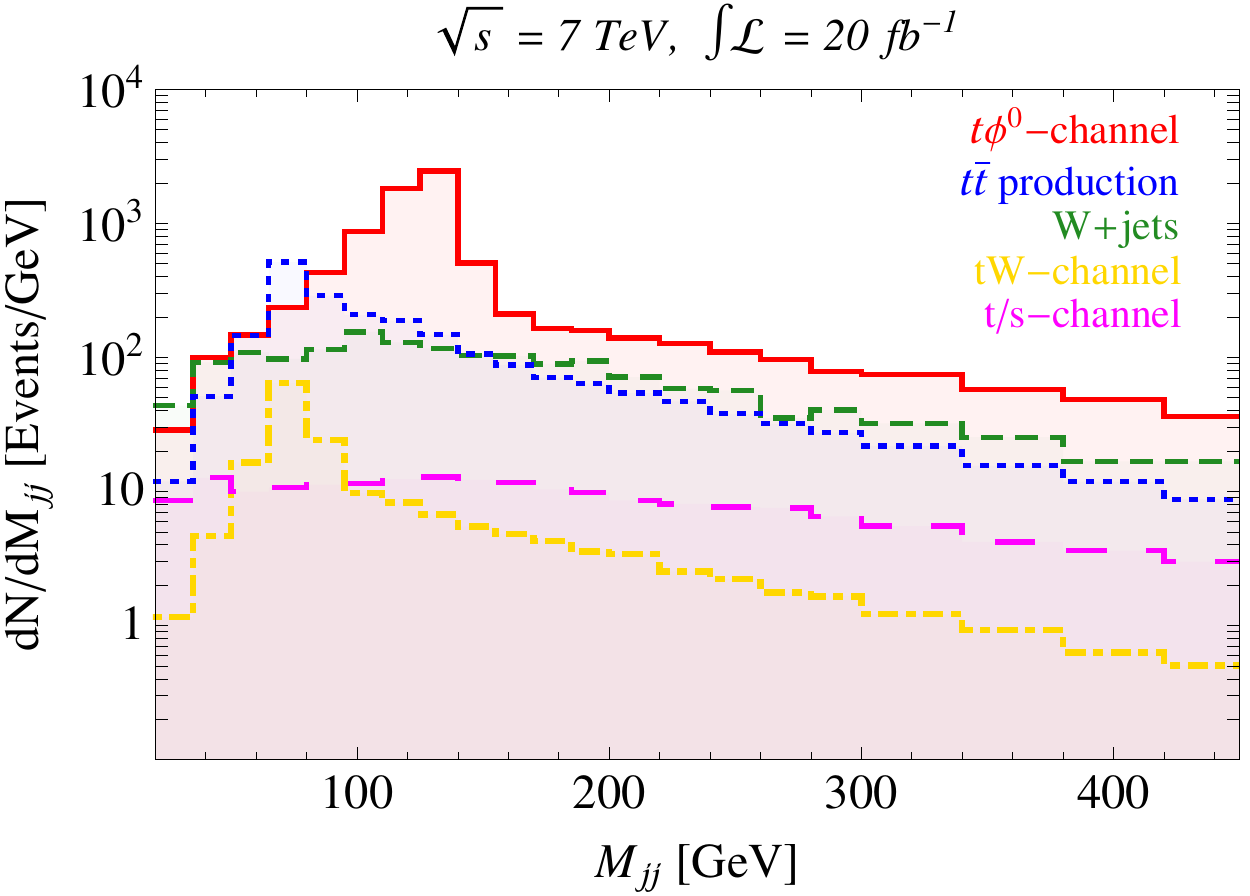}
    \end{minipage}\hspace{2 cm}
  \begin{minipage}{0.4\textwidth}
  \centering
    \includegraphics[scale=0.65]{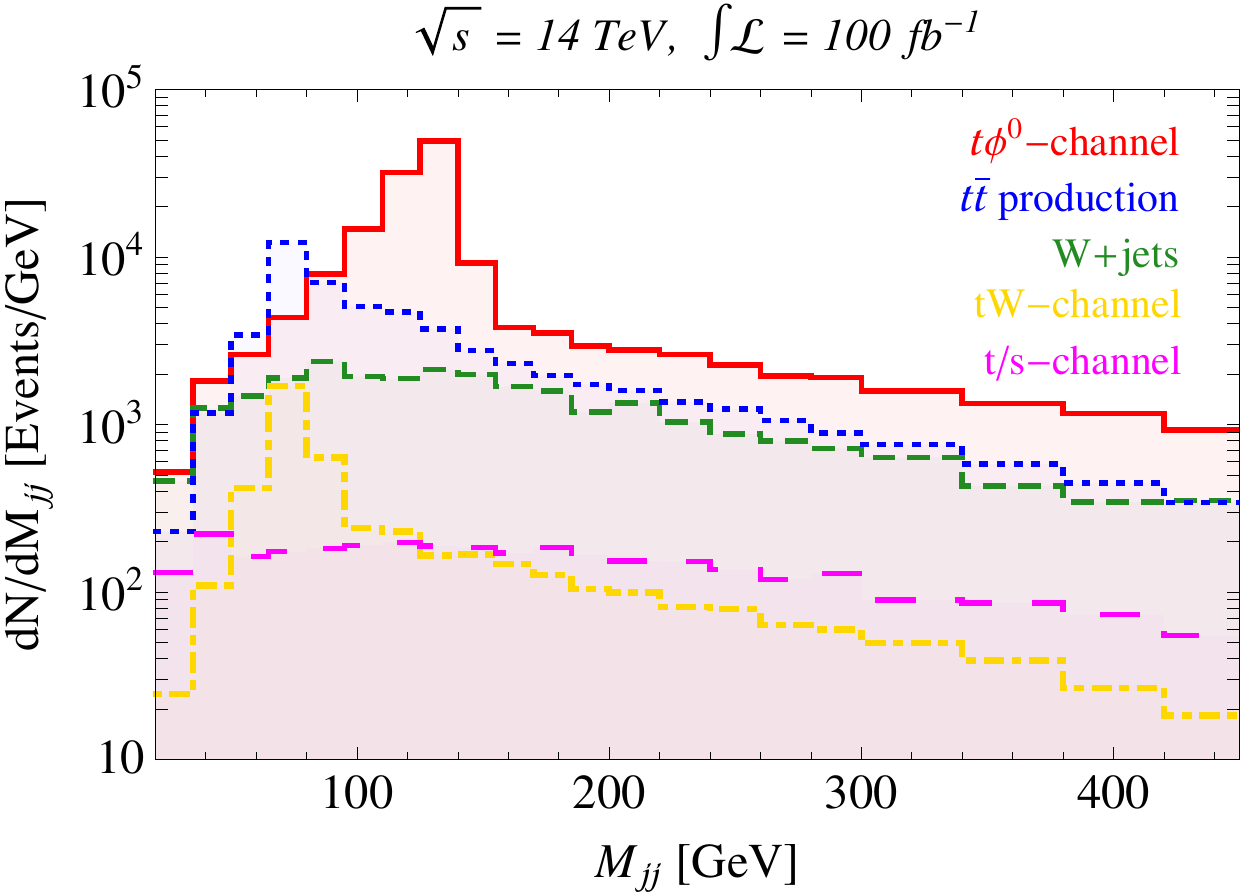}
    \end{minipage}
 \caption{\emph{\textit{Number of generated events as a function of the invariant mass of the two light jets. We consider $\sqrt{s}=7$ TeV, $L=20$ fb$^{-1}$ (left panel) and $\sqrt{s}=14$ TeV, $L=100$ fb$^{-1}$ (right panel).  We show the correspondent distributions considering the signal in the t$\phi^0$-channel (solid red line), the top pairs production (dotted blue line), the $W$+jets channel (green line with small dashes), the tW-channel (yellow dot-dashed line), and the t/s-channel (fuchsia line with large dashes). In both cases a bump around $140$ GeV, corresponding to the mass of the neutral component $\phi^0$, appears on the top of the SM background.}}}
 \label{fig:Milestone}
\end{figure}
We focus on the mass invariant $M_{jj}$ of the two non b-tagged jets with highest $p_T$. This observable, in fact, allows us to reconstruct  the mass of the decaying $\phi^0$ particle leading to a clear footprint of its existence. Our results are shown in Fig. \ref{fig:Milestone} where we present the histograms describing the distribution of events as a function of $M_{jj}$ for $\sqrt{s}=7$ TeV, $L=20$ fb$^{-1}$ and $\sqrt{s}=14$ TeV, $L=100$, respectively. In both cases the $\phi^0$ signal, thanks to its large cross section and  the efficiency of the cuts on the $W$+jets channel, clearly appears around $140$ GeV as a bump over the background distribution.\\
Before switching to $h\to \gamma\gamma$, we would like to make few comments. First, from Tab. \ref{tab:CrossSectionExample}, the cross sections of $ug\to t\phi^0$ might be too large to be consistent with the measurement of the single top quark production in Ref.~\cite{Chatrchyan:2011vp}. Due to the different topology of final states, however, events from t$\phi^0$-channel with 2 light jets probably will not pass the event selection aimed at one light jet. Furthermore, even with $100\%$ acceptance, the total single top quark cross section including SM and NP is still within one $\sigma$ from the measured value after re-scaling mentioned above. Second, low-energy precision tests of parity violation can put stringent constraints on the scalar doublet~\cite{Gresham:2012wc}. In fact, our benchmark point is in significant tension with those bounds but still within $3\sigma$. The semi-leptonic single top production suggested in this paper can serve as an alternative to constrain this type of models which can be achieved at the LHC without resorting to low-energy experiments.

\section{The Higgs decays $h\to \gamma\gamma$, $h\to \gamma Z$}
\label{sec:AFB}

\begin{figure}[!htb!]
  \begin{minipage}{0.4\textwidth}
   \centering
   \includegraphics[scale=0.6]{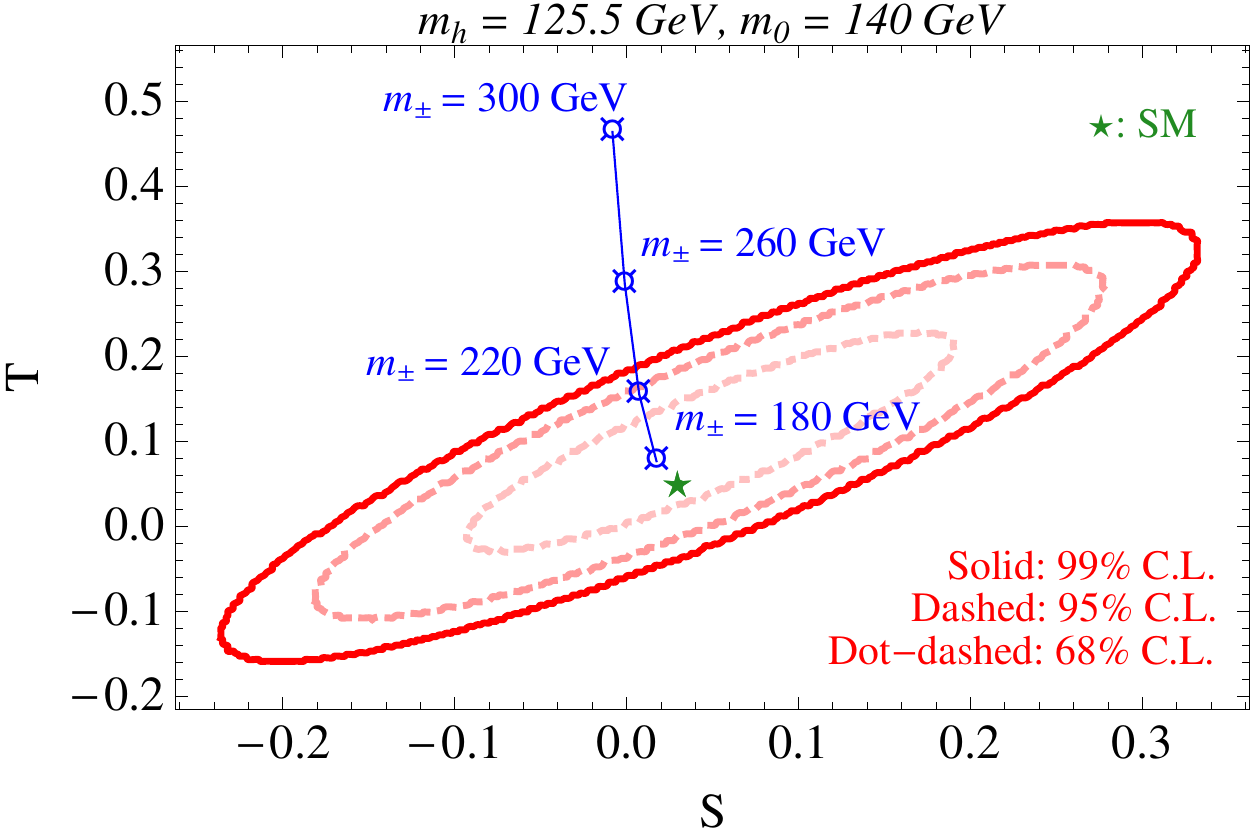}
    \end{minipage}\hspace{1.5 cm}
   \begin{minipage}{0.4\textwidth}
    \centering
    \includegraphics[scale=0.65]{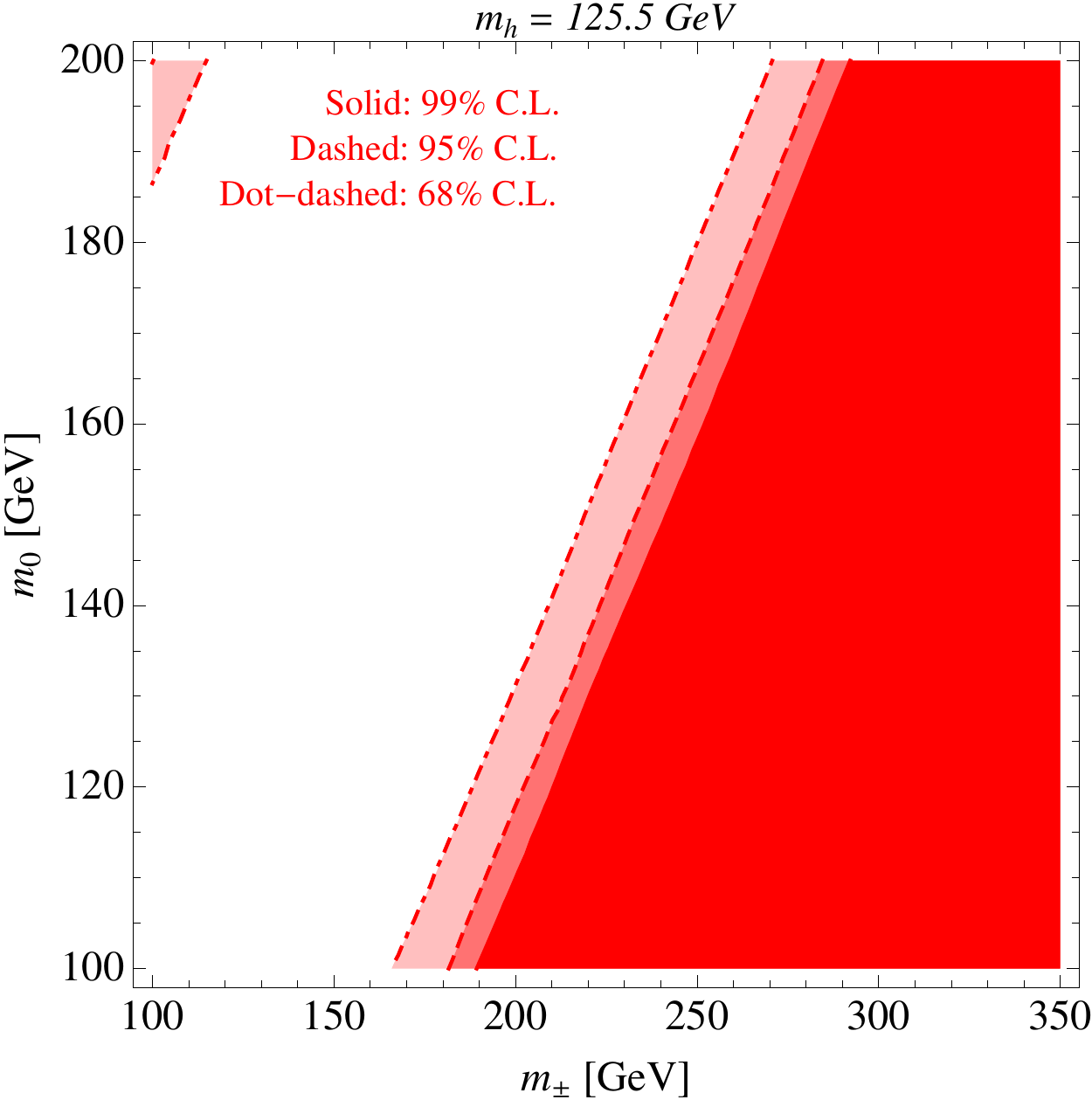}
    \end{minipage}
 \caption{\emph{{\underline{Left panel}}: oblique parameters  $S=S_{\rm SM}+\Delta S_{\phi}$ and  $T=T_{\rm SM}+\Delta T_{\phi}$ in the presence of the extra scalar doublet $\phi$ in addition to the SM particle spectrum with $m_h=125.5$ GeV (green star).
 In Eqs. (\ref{eq:DeltaS},\ref{eq:DeltaT}) we report the correspondent analytical expressions for the deviations $\Delta S_{\phi}$ and $\Delta T_{\phi}$ while we choose from \cite{GFitter} $S_{\rm SM}=0.03$ and $T_{\rm SM}=0.05$. We focus on the value $m_0=140$ GeV for the neutral component $\phi^0$. {\underline{Right panel}}: exclusion plot in the plane $(m_{\pm},m_0)$ obtained considering the constraint on the oblique $S$ and $T$ parameters provided by the electroweak precision observables.}} \label{fig:EllisseST}
 \end{figure}

Let us now turn to discuss the possible implications on the loop-induced Higgs decays $h\to \gamma\gamma$ and $h\to \gamma Z$. In the SM these radiative decays $h\to \gamma\gamma$ and $h\to \gamma Z$ are described by the following widths \cite{Ellis:1975ap}
\begin{eqnarray}
\Gamma_{\rm SM}(h\to \gamma\gamma)&=&\frac{\alpha^2 m_h^3}{256 \pi^3 v^2}
\left|\mathcal{A}_{\rm SM}^{\gamma\gamma}
\right|^2~,
\\
\Gamma_{\rm SM}(h\to \gamma Z)&=&\frac{\alpha^2 m_h^3(1-m_Z^2/m_h^2)^3}{128\pi^3v^2}\left|
\mathcal{A}_{\rm SM}^{\gamma Z}
\right|^2~,
\end{eqnarray}
where $\mathcal{A}_{\rm SM}^{\gamma\gamma}$ and $\mathcal{A}_{\rm SM}^{\gamma Z}$
encompass the structure of the loop functions from the contributions of the spin-$1$ gauge boson $W^{\pm}$ and the spin-$1/2$ top quark, namely
\begin{eqnarray}
\mathcal{A}_{\rm SM}^{\gamma\gamma}&=&
A_1^{\gamma\gamma}\left(\frac{4m_W^2}{m_h^2}\right)+N_CQ_t^2A_{1/2}^{\gamma\gamma}\left(\frac{4m_t^2}{m_h^2}\right)~,
\\
\mathcal{A}_{\rm SM}^{\gamma Z}&=&\frac{c_W}{s_W}\,A_1^{\gamma Z}\left(\frac{4m_W^2}{m_h^2},\frac{4m_W^2}{m_Z^2}\right)+
\frac{N_CQ_t(1/2T_3-Q_ts_W^2)}{s_Wc_W}A^{\gamma Z}_{1/2}\left(\frac{4m_W^2}{m_h^2},\frac{4m_W^2}{m_Z^2}\right)~,
\end{eqnarray}
where for the top quark $N_C=3$, $Q_t=2/3$, $T_3=1/2$. For convenience, we collect in Appendix \ref{app:LoopFunctions} all the loop functions used throughout this Section. \\
After the discovery \cite{Giannotti,:2012gk,Incandela,:2012gu} of a new particle with properties remarkably similar to those of the Higgs boson  in the SM, an accurate  measurement  of its couplings has become of vital importance in order to truly understand the nature of the EWSB. In particular a first glimpse \cite{Giardino:2012dp} to the experimental data shows an excess in the di-photon channel, confirming results from the 2011 data \cite{ATLAS:2012ae,Giardino:2012ww}: the branching ratio of the Higgs boson into two photons seems to be around $1.5$ larger with respect to its SM value. Although it is too premature to come to a conclusion as pointed out in \cite{Baglio:2012et}, it is worthwhile to give a closer look to this experimental result; in principle, any extra charged massive particle couplings to the Higgs boson can change the di-photon decay of the Higgs  (see, for instance, \cite{calderone}) with respect to the SM prediction. Concerning the $h\to \gamma Z$ decay channel, on the other hand, a first analysis of the experimental data will be soon presented. As a result, it is natural to investigate if the charged component $\phi^{\pm}$ of the isospin scalar doublet in Eq. (\ref{eq:ScalarDoublet}) can affect considerably both $h\to \gamma\gamma$ and $h\to \gamma Z$. From the mirror symmetry discussed in Appendix \ref{app:MirrorSymmetry}, we have the scalar potential involving both the Higgs doublet, $H$, and the new scalar doublet, $\phi$,
\begin{equation}\label{eq:ScalarPotential}
V=\mu_1^2|H|^2+\mu_2^2|\phi|^2+\lambda_1|H|^4+\lambda_2|\phi|^4
+\lambda_3|H|^2|\phi|^2+\lambda_4|H^{\dag}\phi|^2+\left[\frac{\lambda_5}{2}(H^{\dag}\phi)^2+h.c.\right]~.
\end{equation}
The contribution of such a scalar spin-0 particle to these decays, that have been computed in \cite{Djouadi:2005gj}, can be quantified by,
\begin{equation}\label{eq:ratioR}
R_{XX}\equiv \frac{\sigma(pp\to h)}{\sigma_{\rm SM}(pp\to h)}\frac{\Gamma(h\to XX)}{\Gamma_{\rm SM}(h\to XX)}.
\end{equation}
Under the assumption that the Higgs boson production is not significantly modified by the existence of the new scalar double, Eq. (\ref{eq:ratioR}) becomes,
\begin{equation}\label{eq:Ratio2}
R_{\gamma\gamma}=\frac{\Gamma(h\to \gamma\gamma)}{\Gamma_{\rm SM}(h\to \gamma\gamma)}~,~~~~~~
R_{\gamma Z}=\frac{\Gamma(h\to \gamma Z)}{\Gamma_{\rm SM}(h\to \gamma\gamma)}~,
\end{equation}
and we have explicitly the following  factors
\begin{eqnarray}
R_{\gamma\gamma}&\equiv&\left|
1+\frac{\lambda_3v^2}{2m_{\pm}^2}\frac{A_0^{\gamma\gamma}\left(\frac{4m_{\pm}^2}{m_h^2}\right)}{\left[A_1^{\gamma\gamma}\left(\frac{4m_W^2}{m_h^2}\right)+N_CQ_t^2A_{1/2}^{\gamma\gamma}\left(\frac{4m_t^2}{m_h^2}\right)\right]}
\right|^2~,\label{eq:RGG}\\
R_{\gamma Z}&\equiv&\left|
1+\frac{\lambda_3v^2}{2m_{\pm}^2}\frac{\frac{(1-2s_W^2)}{s_Wc_W}A_0^{\gamma Z}\left(\frac{4m_{\pm}^2}{m_h^2},\frac{4m_{\pm}^2}{m_Z^2}\right)}{\left[
\frac{c_W}{s_W}\,A_1^{\gamma Z}\left(\frac{4m_W^2}{m_h^2},\frac{4m_W^2}{m_Z^2}\right)+
\frac{N_CQ_t(1/2T_3-Q_ts_W^2)}{s_Wc_W}A^{\gamma Z}_{1/2}\left(\frac{4m_W^2}{m_h^2},\frac{4m_W^2}{m_Z^2}\right)\right]
}
\right|^2~,\label{eq:RGZ}
\end{eqnarray}
where $A_{0}^{\gamma\gamma}$ and $A_0^{\gamma Z}$ are the loop functions describing the contributions of the charged scalar particle $\phi^{\pm}$ with mass $m_{\pm}$.  The rates in Eqs. (\ref{eq:RGG},\ref{eq:RGZ}) are determined by the values of the mass $m_{\pm}$ and the coupling $\lambda_3$.
Based on the decoupling theorem \cite{Appelquist:1974tg}, the larger the mass $m_{\pm}$, the smaller the deviation from the SM rates. One might infer that the contribution of $\phi$ to the top FB asymmetry is independent of that to $h\to \gamma\gamma$ based on the fact they are controlled by different coupling constants ($\lambda$ and $\lambda_3$, respectively) and different masses ($m_0$ and $m_{\pm}$, respevtively\footnote{In principle, they could be different because of EWSB although $\phi^0$ and $\phi^{\pm}$ are in the $SU(2)_L$ doublet.}).  The key, however, to realize the connection between these two observables is to study constraints from the electroweak precision tests, characterized by the oblique $S$ and $T$ parameters,
\begin{eqnarray}
\frac{\alpha S}{4 s_W^2 c_W^2}&=&\Pi^{\prime}_{ZZ}(0)-\Pi^{\prime}_{\gamma\gamma}(0)-\frac{(1-2s_W^2)}{s_W c_W}\Pi_{Z\gamma}^{\prime}(0)~,\label{eq:S}\\
\alpha T&=&\frac{1}{m_W^2}\left[\Pi_{WW}(0)-c_W^2\Pi_{ZZ}(0)\right]~,\label{eq:T}
\end{eqnarray}
where the form factors in Eqs. (\ref{eq:S},\ref{eq:T}) are extracted from the vacuum polarization amplitudes of the electroweak gauge bosons as follows
\begin{equation}
i\Pi_{VV}^{\mu\nu}(p)=i\Pi_{VV}(p^2)g^{\mu\nu}+p^{\mu}p^{\nu}\,{\rm terms}~,~~~V=W,Z,\gamma~,
\end{equation}
with $\Pi_{VV}^{\prime}(0)\equiv \left.d\Pi_{VV}(p^2)/dp^2\right|_{p^2=0}$. $\phi$  modifies the values of $S$ and $T$ with respect to their SM ones \cite{GFitter}. The crucial point is that the resulting $\Delta S_{\phi}$ and $\Delta T_{\phi}$ corrections depend on the mass splitting between the neutral and the charge components $\phi^0$ and $\phi^{\pm}$ \cite{Blum:2011fa,Li:1992dt}; we find
\begin{eqnarray}
\Delta S_{\phi}&=&\frac{1}{12\pi}\ln\frac{m_0^2}{m_{\pm}^2}~,\label{eq:DeltaS}\\
\Delta T_{\phi}&=&\frac{1}{16\pi m_W^2 s_W^2}
\left[
m_0^2+m_{\pm}^2+\frac{2m_0^2m_{\pm}^2}{(m_0^2-m_{\pm}^2)}
\ln\frac{m_{\pm}^2}{m_0^2}
\right]~;\label{eq:DeltaT}
\end{eqnarray}
from Eqs. (\ref{eq:DeltaS},\ref{eq:DeltaT}) it follows that, for a fixed value of $m_0$,
the mass of the charged component $m_{\pm}$ is bounded  by the electroweak precision data.\footnote{Notice that in \cite{Wang:2012zv} the authors consider the simplified limit $m_0=m_{\pm}$. In this limit $\Delta S_{\phi}=\Delta T_{\phi}=0$, and the correlation provided by the oblique parameters disappears.} This situation is shown in the left panel of Fig. \ref{fig:EllisseST} for the representative value $m_0=140$ GeV: at 95\% CL a mass $m_{\pm}$ larger than $220$ GeV  is in conflict with the bound imposed by the correspondent experimental ellipse; a broader view on this constraint is shown in the right panel of Fig. \ref{fig:EllisseST}, where we present the correspondent exclusion plot in the plane $(m_{\pm},m_0)$. The oblique $S$ and $T$ parameters strongly constrain the mass splitting between the two component of the doublet; in particular
for the value $m_0\simeq 140$ GeV suggested by the Tevatron top FB asymmetry, we find that only small values $m_{\pm}\simeq 100-200$ GeV are allowed, that fall exactly into the correct ballpark needed to affect considerably $h\to \gamma\gamma$ and $h\to Z\gamma$.\\

\begin{figure}[!htb!]
    \centering
   \includegraphics[scale=0.7]{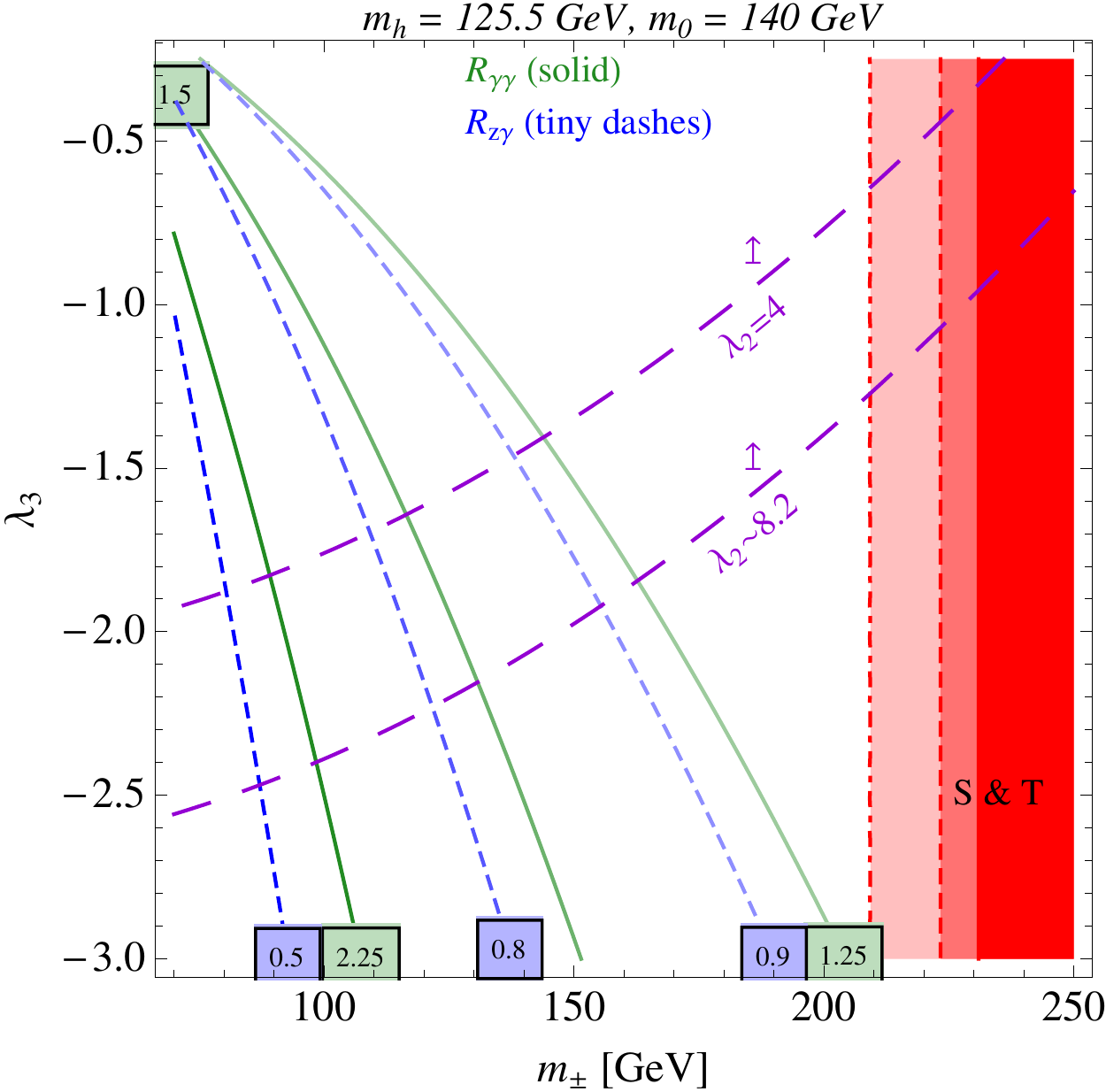}
   \caption{\textit{
   Contours of the factors $R_{\gamma\gamma}$ (green solid lines) and $R_{Z\gamma}$ (blue lines with tiny dashes) in the existence of the extra scalar doublet $\phi$. We show in the same plane $(m_{\pm},\lambda_3)$ the constraints provided by the $S$ and $T$ parameters (red bands) and by the arguments of perturbativity, vacuum stability and unitarity (purple lines with large dashes) summarized in Eqs. (\ref{eq:Constraint1},\ref{eq:Constraint2}).  In particular we show this bound in correspondence of two representative values for $\lambda_2$, namely $\lambda_2=4$ and $\lambda_2=8\pi/3-m_h^2/2v^2\approx 8.2$, while a purple arrow points towards the allowed region.}}\label{fig:Correlazione}
 \end{figure}

Concerning the coupling $\lambda_3$,  first notice that in order to obtain an enhancement in the di-photon rate $R_{\gamma\gamma}$ we need negative values $\lambda_{3}<0$ so as to interfere constructively with the SM $W^{\pm}$ loop. On the other hand, as explained in Appendix \ref{app:MirrorSymmetry}, it is possible to constrain its value requiring  perturbativity, vacuum stability and unitarity for the scalar potential. We find,
\begin{eqnarray}
&& \lambda_3> -\frac{2(m_0^2-m_{\pm}^2)}{v^2}-2\sqrt{\frac{m_h^2\lambda_2}{2v^2}}~, \label{eq:Constraint1}   \\
   &&      0 < \lambda_2 \leqslant \frac{8\pi}{3}-\frac{m_h^2}{2v^2}\approx 8.2 ~;\label{eq:Constraint2}
\end{eqnarray}
for a fixed value of $\lambda_2$ in the range $0<\lambda_2 \lesssim 8.2$ Eq. (\ref{eq:Constraint1}) results in a lower bound for the coupling $\lambda_3$. We show this bound in Fig. \ref{fig:Correlazione} as a function of $m_{\pm}$ considering two representative values for $\lambda_2$, namely $\lambda_2=4$ and $\lambda_2=8\pi/3-m_h^2/2v^2\approx 8.2$.
Even for the maximum  value of $\lambda_2$ allowed by unitarity, we find that large negative values for the coupling $\lambda_3$ are disfavored.\\
  All in all, we show in Fig. \ref{fig:Correlazione} the contours of constant enhancement factors $R_{\gamma\gamma}$ and $R_{Z\gamma}$, defined in Eq. (\ref{eq:Ratio2}), in the plane $(m_{\pm},\lambda_3)$ considering the value $m_0=140$ GeV as the best-suited one responsible for the top asymmetries; we also include the constraints from the oblique parameters $S$ and $T$, and those from the requirement of perturbativity, vacuum stability and unitarity from the scalar potential. An enhancement $R_{\gamma\gamma}\simeq 1.25-2.25$ in the di-photon channel is possible in the region $80\lesssim m_{\pm}\lesssim 130$ GeV and $-1.5\lesssim \lambda_3 \lesssim -0.5$. In this region, the model predicts a moderate suppression on $H\to Z\gamma$ with $0.6\lesssim R_{Z\gamma}\lesssim 0.9$.\\
Before coming to conclusions, we would like to comment on some collider constraints. All four LEP experiments have searched for light charged scalars  in the context of the MSSM, and their combined results put a lower bound on the mass of the sleptons of the order of  $100$ GeV \cite{LEPsleptons}. Notice, however, that the bound does not strictly apply here. Sleptons, in fact, decay predominantly into their SM partners and the lightest neutralino, leading to an experimental signature characterized by a pair of oppositely-charged leptons,
accompanied with missing energy carried away by the two undetected neutralinos. On the contrary in our model the fermionic decay of the charged scalar $\phi^{\pm}$ involves only light quarks, as described by the Lagrangian in Eq. (\ref{eq:PhenoLagrangian}), thus resulting in a different final state.

\section{Conclusive remarks}
\label{sec:Conclusions}

In this paper, we try to explain and relate with an extra scalar doublet two recent experimental results: the top FB asymmetry measured at the Tevatron, and the excess in the di-photon Higgs decay rate at the LHC. The aim of this paper is twofold.
\begin{enumerate}
\item Following the analysis proposed in \cite{Blum:2011fa}, we have reconsidered the role of the scalar doublet in the context of the top FB asymmetry at the Tevatron, and the top charge asymmetry at the LHC. We focus on light masses for the neutral component of the doublet, identifying a representative value $m_0=140$ GeV as our best-suited candidate. The charged component can simultaneously modify the loop-induced Higgs decays, $h\to \gamma\gamma$ and $h\to Z\gamma$. Allowing for different values between the masses of these two components, we have shown how the constraints from the electroweak precision tests provide a correlation for the mass splitting, thus connecting these two experimental observables.
\item Looking for complementary experimental evidences, we have considered the semi-leptonic single top production in the tW-channel at the LHC. We have shown that the interactions needed to explain the FB asymmetry manifest as a bump in the mass invariant distribution of the two light jets. Hence, this theoretical prediction addresses the necessity to pursue at the LHC a dedicated experimental analysis in this particular channel.
\end{enumerate}
If these two anomalies at the Tevatron and LHC turn out to be true, then the  perspective proposed in this paper could represent  an interesting and falsifiable phenomenological scenario. On the other hand, the concrete realization of the origin of the scalar doublet and its special structure on Yukawa interactions goes well beyond the phenomenological purpose of this paper, and it will be investigated in a forthcoming publication.

\subsection*{Aknowledgement}

We especially thank Denis Comelli  for encouragement and advices when the first idea of this paper was born. We also thank Arian Abrahantes and Tim M. P. Tait for useful suggestions about MadGraph\,5, and Nicola Orlando for interesting discussions about the LHC physics. The work of AU is supported by the ERC Advanced Grant n$^{\circ}$ $267985$, Electroweak Symmetry Breaking, Flavour and Dark Matter: One Solution for Three Mysteries" (DaMeSyFla).

\appendix
\section{Scalar doublet from a mirror symmetry}\label{app:MirrorSymmetry}
The desired $SU(2)_L$ doublet can arise from the Two-Higgs-Doublet Model (2HDM), $\Phi_1$ and $\Phi_2$ with a mirror symmetry. The mirror symmetry  renders the Lagrangian invariant under the interchange between $\Phi_1$ and $\Phi_2$, i.e., $\mathcal{L}(\Phi_1,\Phi_2)=\mathcal{L}(\Phi_2,\Phi_1)$. The most general renormalizable Higgs potential under the mirror symmetry can be written as,

\bea
V&=&m^{\prime\,2}(\Phi_1^{\dag}\Phi_1+\Phi_2^{\dag}\Phi_2)-m_{12}^{\prime\,2}(\Phi_1^{\dag}\Phi_2+\Phi_2^{\dag}\Phi_1) + y_1 \left[ (\Phi_1^{\dag}\Phi_1)^2+(\Phi_2^{\dag}\Phi_2)^2\right] \notag
\\ && +y_2 (\Phi_1^{\dag}\Phi_1)(\Phi_2^{\dag}\Phi_2) +y_3 (\Phi_1^{\dag}\Phi_2)(\Phi_2^{\dag}\Phi_1)
+y_4 \left[   (\Phi_1^{\dag}\Phi_2)^2  + (\Phi_2^{\dag}\Phi_1)^2   \right]  \notag\\
 &&+y_5 \left[ (\Phi_1^{\dag}\Phi_2)(\Phi_1^{\dag}\Phi_1)+(\Phi_2^{\dag}\Phi_1)(\Phi_2^{\dag}\Phi_2)  \right]
+y^*_5 \left[ (\Phi_2^{\dag}\Phi_1)(\Phi_1^{\dag}\Phi_1)+(\Phi_1^{\dag}\Phi_2)(\Phi_2^{\dag}\Phi_2)  \right] ,
\eea
where except for $y_5$, all parameters are real. After EWSB, $\Phi_1$ and $\Phi_2$ obtain an equal vacuum expectation value (vev) due to the mirror symmetry. We can rotate the $\Phi_1-\Phi_2$ basis into that of $H-\phi$, where $H=\frac{1}{\sqrt{2}}(\Phi_1 + \Phi_2)$ is the SM Higgs doublet with $v\equiv \sqrt{2}\langle H\rangle=246$ GeV, and $\phi=\frac{1}{\sqrt{2}}(-\Phi_1 + \Phi_2)$ is the extra $SU(2)_L$ doublet, with $\langle\phi\rangle=0$. Now, the Higgs potential can be expressed in terms of $H$ and $\phi$.

\begin{equation}\label{eq:ScalarPotential}
V=\mu_1^2|H|^2+\mu_2^2|\phi|^2+\lambda_1|H|^4+\lambda_2|\phi|^4
+\lambda_3|H|^2|\phi|^2+\lambda_4|H^{\dag}\phi|^2+\left[\frac{\lambda_5}{2}(H^{\dag}\phi)^2+h.c.\right]~.
\end{equation}
For simplicity, $\lambda_5$  is set to zero from now on to eliminate a mass splitting between the two components of the complex scalar, $\phi^{0}$.
Note that there are no terms with odd powers of $\phi$ due to $\phi \rightarrow -\phi$ ($H \rightarrow H$) under the mirror symmetry that in turn forbids the unwanted mixing between $H$ and $\phi$.
From this point of view, the 2HDM with the mirror symmetry is basically equivalent to the Inert Higgs Doublet Model (IHDM)~\cite{Deshpande:1977rw}. A closer look to the scalar potential in Eq. (\ref{eq:ScalarPotential}) allows us to highlight the following properties.
\begin{enumerate}[i.]
\item {\underline{\textit{Mass spectrum}}}. After EWSB the Higgs mass and the masses of the two components in the doublet $\phi$ can be expressed in terms of the coefficient of the scalar potential in Eq. (\ref{eq:ScalarPotential}) as follows,
\begin{eqnarray}
m_h^2&=& -2\mu_1^2=2\lambda_1v^2~,\label{eq:HiggsMass}\\
m_{\pm}^2&=& \mu_2^2+\lambda_3v^2/2~,\\
m_0^{2}&=& \mu_2^2 +(\lambda_3+\lambda_4)v^2/2~.
\end{eqnarray}

\item {\underline{\textit{Perturbativity, vacuum stability and unitarity}}}. The constraints on the Higgs potential derived in Refs.~\cite{Arhrib:2012ia,Lee:1977eg} apply here, which amount to
\bea
&&|\lambda_i| \leq 8\pi~, \\
&&\lambda_{1,2} >0~,~~  \lambda_3+\lambda_4+\sqrt{\lambda_1\lambda_2}>0~,~~ \lambda_3 +2 \sqrt{\lambda_1\lambda_2}>0~,\label{eq:VS}\\
&&\lambda_1 + \lambda_2\leq  8\pi/3~.\label{eq:Unitarity}
\eea
\end{enumerate}
Notice that combining the mass spectrum with Eqs. (\ref{eq:VS},\ref{eq:Unitarity}) we find the following bounds
\begin{eqnarray}
&& \lambda_3> -\frac{2(m_0^2-m_{\pm}^2)}{v^2}-2\sqrt{\frac{m_h^2\lambda_2}{2v^2}}~, \label{eq:Constraint1bis}   \\
   &&      0 < \lambda_2 \leqslant \frac{8\pi}{3}-\frac{m_h^2}{2v^2}\approx 0.82 ~,\label{eq:Constraint2bis}
\end{eqnarray}
where we assume throughout this work $m_h=125.5$ GeV. As shown in Section \ref{sec:AFB}, these bounds further constrain the radiative electroweak Higgs decays in light of the results derived in Sections \ref{sec:FBTevatron}, \ref{sec:FBLHC} from the top asymmetries.\\
Now, we are in a position to study the interplay between $\phi$ and fermions. If the mirror symmetry holds in Yukawa couplings, only $H$ couples to fermions. We, however, need to break the symmetry so that $\phi$ also talks to fermions without affecting $H$-fermion couplings. It can be achieved  if breaking is ``maximal", i.e., fermions couple to $\Phi_1$ and $\Phi_2$ with an opposite sign such that $H$-fermion interactions are untouched and $\phi$ obtains Yukawa couplings\footnote{In general, we can have $c_1 \bar{L}_f \Phi_1 f + c_2 \bar{L}_f\Phi_2 f$ and $c_1\neq c_2$. In this case, $c_1-c_2$ will manifest in the $\phi$ Yukawa couplings and $c_1+c_2$ in those of $H$. It, however, modifies the CKM matrix and in turn might induce the mixing between $H$ and $\phi$. Therefore, we constrain ourself to maximal breaking.}. To be more precise, we have,
\be
\mathcal{L} \supset -c \bar{L}_f \Phi_1 f + c \bar{L}_f\Phi_2 f +h.c.~=\sqrt{2}\bar{L}_f\phi f+h.c.~,
\ee
where $L_f$ and $f$ are the $SU(2)_L$ doublet and singlet fermions. This kind of breaking is different from the breaking mechanism on $Z_2$ symmetry in the context of IHDM in Refs.~\cite{Wang:2012zv,Cao:2011yt} in the following way. First, if $Z_2$ is imposed, i.e., $H\rightarrow H$ and $\phi\rightarrow-\phi$, then for fermions coupling to $\phi$, some fermion(s) has to carry the $Z_2$ charge which prevents a mass term from the Higgs mechanism because of the even parity of $H$. A soft breakdown of the $Z_2$-symmetry, through a non-vanishing $\mu_{12}^2 H^{\dag}\phi$ term, would be enough to provide a mass to light fermions. In the mirror symmetry, Yukawa couplings of $H$ and fermions are uninfluenced by virtue of ``maximal" breaking. Second, at one loop, once $Z_2$ is broken, the $\mu_{12}^2 H^{\dag}\phi$ term will be generated by radiative corrections which might result in a large vev for $\phi$ and large mixing between $H$ and $\phi$. Having $\phi$ couple to the first generation can ease the tension in addition to a cancelation between the tree-level and loop-induced contributions. For the mirror symmetry, as long as $\phi$ couples to fermions with different generations, $H-\phi$ mixing will not arise at one loop because of vanishing off-diagonal terms in $H$ Yukawa couplings in the mass basis\footnote{In fact, Eq.(\ref{eq:PhenoLagrangian}) on which this paper based induces the mixing between $H$ and $\phi$ but it is highly suppressed by the small u-quark Yukawa coupling and small $|V_{ub}|$.}.

Finally, we would like to briefly comment on the dark matter. The doublet $\phi$ is unstable and decays into quarks due to breaking of the mirror symmetry, which makes 2HDM not have a viable dark matter candidate. It can be overcome simply by the introduction of the scalar gauge singlet under certain $Z_2$ symmetry. Besides, from the spirit of the mirror symmetry, we can add one more $SU(2)_L$ doublet, $\Phi_3$, so that the Lagrangian is invariant under interchange of $\Phi_i$'s $(i=1,2,3)$, and break this symmetry with a term, $c (\bar{L}_f \Phi_1 f +  \bar{L}_f\Phi_2 f - 2 \bar{L}_f\Phi_3 f)$. In this case, the neutral component of $\Phi_1 - \Phi_2$ can be the dark matter.

\section{Loop functions}\label{app:LoopFunctions}
In this Appendix we collect the loop functions used throughout this paper.
\subsection{Di-photon Higgs decay}
\begin{eqnarray}
A_{0}^{\gamma\gamma}(x)&=& -x^2\left[\frac{1}{x}-f\left(\frac{1}{x}\right)\right]~,  \\
A_{1/2}^{\gamma\gamma}(x)&=& 2x^2\left[\frac{1}{x}+\left(\frac{1}{x}-1\right)f\left(\frac{1}{x}\right)\right]~, \\
A_{1}^{\gamma\gamma}(x)&=& -x^2\left[\frac{2}{x^2}+\frac{3}{x}+3\left(\frac{2}{x}-1\right)f\left(\frac{1}{x}\right)\right]~, \\
\end{eqnarray}
where
\begin{equation}
f(x)=\arcsin^2\sqrt{x}~.
\end{equation}

\subsection{$Z\gamma$ Higgs decay}
\begin{eqnarray}
A_{0}^{\gamma Z}(x,y)&=& \mathcal{I}_{1}(x,y)~,  \\
A_{1/2}^{\gamma Z}(x,y)&=& 4\left[ \mathcal{I}_{1}(x,y)- \mathcal{I}_{2}(x,y)\right]~, \\
A_{1}^{\gamma Z}(x,y)&=&
4\left(3-\frac{s_W^2}{c_W^2}\right)\mathcal{I}_2(x,y)+
\left[\left(1+\frac{2}{x}\right)\frac{s_W^2}{c_W^2}-\left(5+\frac{2}{x}\right)\right]\mathcal{I}_1(x,y)~,
\end{eqnarray}
where
\begin{eqnarray}
\mathcal{I}_{1}(x,y)&=&
\frac{x y}{2(x-y)}+\frac{x^2 y^2}{2 (x-y)^2}\left[f\left(\frac{1}{x}\right)-f\left(\frac{1}{y}\right)\right]+\frac{x^2 y}{(x-y)^2} \left[g\left(\frac{1}{x}\right)-g\left(\frac{1}{y}\right)\right]~,\nonumber\\
 \\
\mathcal{I}_{2}(x,y)&=& -\frac{x y}{2(x-y)}\left[f\left(\frac{1}{x}\right)-f\left(\frac{1}{y}\right)\right]~,
\end{eqnarray}
with
\begin{equation}
g(x)=\sqrt{\frac{1}{x}-1}\arcsin\sqrt{x}~.
\end{equation}

\newpage



\end{document}